\documentclass[twocolumn,aps,prd,nofootinbib,showpacs]{revtex4}
\usepackage{amsmath,graphicx,bm}
\begin{document}

\title{Effects of cold dark matter decoupling \\
  and pair annihilation on cosmological perturbations
}
\author{Edmund Bertschinger}
\affiliation{Department of Physics, MIT Room 37-602A, 77
Massachusetts Ave., Cambridge, MA 02139}
\date{\today}

\begin{abstract}
Weakly interacting massive particles are part of the lepton-photon
plasma in the early universe until kinetic decoupling, after which
time the particles behave like a collisionless gas with nonzero
temperature. The Boltzmann equation for WIMP-lepton collisions is
reduced to a Fokker-Planck equation for the evolution of the WIMP
distribution including scalar density perturbations.  This equation
and the Einstein and fluid equations for the plasma are solved
numerically including the acoustic oscillations of the plasma before
and during kinetic decoupling, the frictional damping occurring
during kinetic decoupling, and the free-streaming damping occurring
afterwards and throughout the radiation-dominated era.  An excellent
approximation reduces the solution to quadratures for the cold dark
matter density and velocity perturbations. The subsequent evolution
is followed through electron pair annihilation and the
radiation-matter transition; analytic solutions are provided for
both large and small scales. For a 100 GeV WIMP with bino-type
interactions, kinetic decoupling occurs at a temperature $T_d=23$
MeV. The transfer function in the matter-dominated era leads to an
abundance of small cold dark matter halos; with a smooth window
function the Press-Schechter mass distribution is $dn/d\ln M\propto
M^{-1/3}$ for $M<10^{-4}\,(T_d/$10 MeV)$^{-3}$ M$_\odot$.
\end{abstract}

\maketitle

\section{Introduction}
\label{sec:intro}

Weakly interacting massive particles (WIMPs) are perhaps the leading
candidate for the cold dark matter (CDM) making up most of the
nonrelativistic mass density of the universe today \cite{bhs}.
Although candidate WIMPs are 10 to 1000 times more massive than
nucleons and have no electromagnetic or color charges, their cosmic
histories share many parallels with nucleons. After their abundances
froze out at $\sim10^{-10}$ the number density of photons, both
WIMPs and nucleons remained thermally coupled to the plasma by
elastic scattering with abundant relativistic particles.  Acoustic
oscillations in the relativistic plasma imprinted oscillations on
both WIMPs and nucleons.  Eventually the plasma released its grip on
both types of particles.  For WIMPs, this event is called kinetic
decoupling; for nucleons, recombination.

During kinetic decoupling, friction between the WIMP gas and
relativistic plasma led to Silk damping of small-scale waves similar
to what happened much later for atomic matter at recombination.
After the respective decoupling periods ended, pressure forces (and
in the case of WIMPs, shear stress) inhibited gravitational
instability on small scales. Still later, both WIMPs and nucleons
played a major role in galaxy formation.

There are, of course, significant differences between the cosmic
evolution of WIMPs and nucleons.  Most evident are the quantitative
differences: because their interactions are so weak, WIMPs decoupled
from the plasma less than one second after the big bang.
Consequently the WIMP acoustic oscillations appear only on a length
scale vastly smaller ($\sim$ parsec scales) than the baryon acoustic
oscillations. The physics of nucleon decoupling, as imprinted in the
galaxy distribution \cite{eisenstein} and in the cosmic microwave
background radiation \cite{wmap3} provides a powerful probe of
cosmic parameters and inflationary cosmology.  If it were possible
to similarly measure fluctuations on the scale of WIMP acoustic
oscillations, we would have a dramatic consistency test of the
cosmological model as well as an astrophysical measurement of WIMP
properties.

One way to constrain the parameters of cold dark matter decoupling
is to measure the mass function of the smallest dark matter clumps
today \cite{hss01,bde03,ghs04}.  Such clumps would be far too
diffuse to host observable concentrations of atomic matter. However,
they might be observable through the products of the very rare
WIMP-WIMP annihilations taking place in the cores of these objects.
Diemand {\it et al.}\ \cite{diemand} proposed that numerous
Earth-mass clumps might survive to the present day and provide a
detectable gamma-ray signal.  The mass and abundance of these clumps
depends on cosmic fluctuation evolution during and after kinetic
decoupling.

WIMPS and nucleons also differ in a qualitative manner which has
important consequences for the evolution of fluctuations through
kinetic decoupling.  After recombination, elastic scattering is
rapid enough for atoms (and the residual free electrons) to behave
as a nearly perfect gas on cosmological scales.  Baryons behave like
a fluid.  WIMPs, however, collide too infrequently to thermalize
after kinetic decoupling.  WIMPs behave like a collisionless gas.
Different approximations to the evolution of this collisionless gas
have led to different results for the small-scale transfer function
of CDM fluctuations \cite{ghs04,ghs05,lz05}.

In the present paper, the transfer functions for CDM fluctuations
are calculated starting from the full Boltzmann equation describing
elastic scattering between WIMPs and the relativistic leptons
present before neutrino decoupling and electron-positron pair
annihilation. Because the momentum transfer per collision with
nonrelativistic WIMPs is small, the Boltzmann equation reduces to
the Fokker-Planck equation describing diffusion in velocity space
caused by elastic scattering, combined with advection and
gravitational forces.  The Fokker-Planck equation fully describes
kinetic decoupling and the evolution of perturbations of any length
scale without approximating the WIMPs either as a perfect fluid or
fully collisionless gas. Although the solution of the perturbed
Fokker-Planck equation is more difficult than the solution of
coupled fluid and collisionless Boltzmann equations, it is both
numerically and analytically tractable (with an excellent
approximation) in the present case.

After kinetic decoupling, two additional events have an effect on
the CDM transfer function.  The first is electron-positron pair
annihilation, which changes the equation of state of the plasma
thereby modifying the evolution of fluctuations.  Although the
effects are small, they can be analytically calculated.  The more
important event is the transition from a radiation- to
matter-dominated universe occurring about $10^5$ years after the big
bang.  If the photons and neutrinos are treated as fluids, it is
possible to get analytic results for the linear evolution all the
way to low redshift which are accurate to a few percent.  With these
results in place, using standard techniques it is straightforward to
estimate the mass function of CDM clumps at high redshift.

\section{Evolution of WIMP perturbations through kinetic decoupling}
\label{sec:silk}

Weakly interacting dark matter particles are described by their
phase space density, which obeys the Boltzmann equation governing
transport by collisions with leptons (during kinetic decoupling
these are just electrons, positrons, and neutrinos).  For
definiteness we will take the WIMP to be the lightest neutralino
$\chi^0$, however the results are easily applied to other WIMP
candidates by modifying the scattering matrix element below.

Let $f_\chi({\bf p})$ and $f_L({\bf p})$ be the proper phase space
densities of neutralinos and ultrarelativistic leptons,
respectively, where ${\bf p}$ is the proper three-momentum in a
local orthonormal frame.  (The spacetime coordinates ${\bf x}$ and
$t$ are suppressed for brevity.) The phase space densities are
normalized so that $\int f({\bf p})\,d^3p$ is the spatial number
density, summed over spin states (we assumed unpolarized spins). One
distribution function suffices for the relativistic leptons because
electroweak interactions maintain local thermal equilibrium at a
temperature $T_L$. Elastic scattering of neutralinos and leptons
cause the neutralino distribution to evolve according to the
Boltzmann equation,
\begin{eqnarray}\label{neuboltz}
  p^\mu_1\partial_\mu f_{1\chi}&=&\int\frac{d^3p_2}{E_2}
    \int\frac{d^3p_3}{E_3}\int\frac{d^3p_4}{E_4}\left\vert
    \frac{{\cal M}}{8\pi}\right\vert^2\nonumber\\
  &&\times\delta^4(p_1+p_2-p_3-p_4)\nonumber\\
  &&\times\left[f_{3\chi}f_{4L}(1-\tilde f_{2L})-f_{1\chi}f_{2L}
    (1-\tilde f_{4L})\right]\ ,\quad\quad
\end{eqnarray}
where ${\cal M}$ is the Lorentz-invariant scattering amplitude,
$f_{1\chi}\equiv f_\chi({\bf p}_1)$ and similarly for the other
distribution functions, $E_i$ is the energy of particle $i$, and
$\tilde f_L\equiv (2\pi\hbar)^3f_L/2$ is the occupation number.
Pauli blocking must be included for the leptons but not for the
neutralinos since the latter have a low density after chemical
decoupling.  Equation (\ref{neuboltz}) is relativistically covariant
but gives only the effects of collisions; the effects of
gravitational perturbations will be added later. Assuming
effectively massless leptons, the matrix element for slepton
exchange is given in Appendix A of \cite{hss01} and may be written
\begin{equation}\label{matelem}
  \vert{\cal M}\vert^2=C\frac{p_{\rm CM}^2}{m_\chi^2}
    \left(1-\frac{t_{\rm M}}{4p_{\rm CM}^2}\right)\ ,
\end{equation}
where
\begin{equation}\label{Cneut}
  C=256\sum_{L}(b_L^4+c_L^4)
  \left(\frac{G_Fm_W^2m_\chi^2}{m_{\tilde L}^2
  -m_\chi^2}\right)^2
\end{equation}
is a dimensionless constant depending on the relevant particle
masses and couplings ($G_{\rm F}$ is the Fermi constant, $b_L$ and
$c_L$ are left and right chiral vertices, and $m_W$, $m_{\tilde L}$,
and $m_\chi$ are respectively the masses of the $W$ boson, the
slepton, and the neutralino; $G_Fm_W^2=0.0754$).  Here we assume
following Ref.\ \cite{hss01} that the neutralino is a pure bino.
Additionally, $p_{\rm CM}$ is the momentum in the center of momentum
frame, and (using metric signature $-+++$) $t_{\rm M}=
-(p_2-p_4)^2=-2p_{\rm CM}^2(1-\cos\theta_{\rm CM})$ is one of the
Mandelstam variables.

For $p_{\rm CM}\sim T_L\ll m_\chi$, $p_{\rm CM}=m_\chi\epsilon
[1-\epsilon+\frac{3}{2} \epsilon^2+O(\epsilon^3)]$ where
$\epsilon\equiv-p_3\cdot p_4/ m_\chi^2$.  In the lab frame, working
to first order in $T_L/m_\chi$, assuming $p_1\sim\sqrt{m_\chi T_L}$,
the collision kinematics gives
\begin{eqnarray}\label{thetacm}
  \frac{1-\cos\theta_{\rm CM}}{1-{\bf n}_2\cdot{\bf n}_4}=
    1+\frac{({\bf n}_2+{\bf n}_4)\cdot({\bf p}_1+{\bf p}_2)}
    {m_\chi}\quad\quad\quad\quad\quad\quad&&\nonumber\\
    +\left(\frac{p_1}{m_\chi}\right)^2\left(\mu_{12}^2
    +\mu_{12}\mu_{14}+\mu_{14}^2-1\right)\ ,\quad&&
\end{eqnarray}
where $\mu_{ij}={\bf n}_i\cdot{\bf n}_j$ and ${\bf n}_i={\bf
p}_i/p_i$.  Another useful relation follows from energy
conservation,
\begin{equation}\label{momratio}
  \frac{p_4}{p_2}=1+\frac{p_1}{m_\chi}(\mu_{14}-\mu_{12})\left(1
    +\frac{p_1}{m_\chi}\mu_{14}\right)+\frac{p_2}{m_\chi}
    (\mu_{24}-1)\ ,\quad
\end{equation}
valid again to first order in $T_L/m_\chi$.

Frequent collisions among the leptons maintain thermal equilibrium.
Assuming negligible chemical potential, for each species of massless
lepton we have
\begin{equation}\label{flep}
  [\tilde f_L({\bf p})]^{-1}=1+\exp\left[\frac{p}{T_L}(1-
    {\bf n}\cdot{\bf v}_L)\right]\ ,
\end{equation}
where ${\bf v}_L$ is the (very small) local lepton fluid velocity
due to cosmological perturbations.  It is easy to check that an
equilibrium solution of (\ref{neuboltz}) is then the
Maxwell-Boltzmann distribution with mean velocity ${\bf v}_L$.

For $T_L\ll m_\chi$ the $f_{3\chi}$ term may be Taylor-expanded in
(\ref{neuboltz}).  After a lengthy calculation using
(\ref{thetacm})--(\ref{flep}), one obtains $p^\mu\partial_\mu
f=m_\chi(df/dt)_c$ (dropping the subscript on $f_\chi$) where the
Boltzmann collision integral becomes the Fokker-Planck operator,
\begin{equation}\label{boltzop}
\left(\frac{df}{dt}\right)_c=\gamma\frac{\partial}
  {\partial{\bf p}}\cdot\left[({\bf p}-m_\chi{\bf v}_L)f
  +m_\chi T_L\frac{\partial f}{\partial{\bf p}}\right]\ ,
\end{equation}
where
\begin{equation}\label{gamma}
  \gamma=\frac{155\pi^3CT_L^6}{6048m_\chi^5}
\end{equation}
is a rate coefficient (in units where $\hbar=c=1$).  Our exact
result for the rate coefficient is larger by a factor 9.9 than the
estimate obtained from Eqs.\ (9) and (12) of Ref.\ \cite{hss01} and
by a factor 3.4 than Eq.\ (17) of Ref. \cite{bde03}.  The rate is
greater than the simple estimates made in previous work because of
the details of the kinematics and the near-cancelation of forward
and inverse rates in (\ref{neuboltz}).  A larger rate coefficient
leads to a lower temperature for kinetic decoupling than previous
estimates.

If we neglect spatial inhomogeneities, the unperturbed phase space
density $f_0(q,\tau)$ depends on both comoving momentum $q=ap$ and
conformal time $\tau$ according to
\begin{equation}\label{boltz0}
  \frac{\partial f_0}{\partial\tau}=\gamma a\frac{\partial}
    {\partial{\bf q}}\cdot\left[{\bf q}f_0+a^2m_\chi T_L
    \frac{\partial f_0}{\partial{\bf q}}\right]\ .
\end{equation}
Amazingly, for any time-dependence of $\gamma$, $a$, and $T_L$ an
exact solution to this Fokker-Planck equation is the
Maxwell-Boltzmann distribution
\begin{equation}\label{maxboltz}
  f_0(q,\tau)=\frac{\exp(-q^2/2\sigma_q^2)}
    {(2\pi\sigma_q^2)^{3/2}}\ ,\ \
    \sigma_q(\tau)=a(\tau)\sqrt{m_\chi T_\chi(\tau)}\ ,
\end{equation}
where the WIMP temperature $T_\chi$ follows from integrating
\begin{equation}\label{tneude}
  \frac{d\ln(a^2T_\chi)}{d\tau}=2\gamma a\left(\frac{T_L}
    {T_\chi}-1\right)\ .
\end{equation}
During adiabatic evolution in the early universe, $T_L\propto
a^{-1}\propto\tau^{-1}$ and the WIMP proper temperature is then
given in terms of the incomplete Gamma function by
\begin{equation}\label{tneu}
  T_\chi(\tau)=T_Ls^{1/4}e^s\,\Gamma({\textstyle\frac{3}{4}},s)\ ,\ \
  s\equiv\frac{1}{2}\gamma a\tau=\frac{\gamma}{2H}\ .
\end{equation}
Equation (\ref{maxboltz}) may be multiplied by any constant,
allowing the comoving number density of WIMPs to be normalized to
its value after freeze-out.

Familiarity with Brownian motion makes it seem natural that the
solution to the Fokker-Planck equation is a Maxwell-Boltzmann
distribution.  However, the lepton temperature $T_L$ and the
momentum transfer rate $\gamma$ are falling with time and
WIMP-WIMP elastic scattering is far too slow to thermalize the
distribution. Even so, collisions with the leptons maintain the
WIMPs in a thermal distribution with a temperature that deviates
increasingly from the lepton temperature throughout kinetic
decoupling.  Once kinetic decoupling is complete the WIMP momenta
redshift as $p\propto a^{-1}$ preserving the Maxwell-Boltzmann
distribution with $T_\chi\propto a^{-2}$.

Long before kinetic decoupling ($s\gg1$),
$T_\chi/T_L=1-1/(4s)+O(s^{-2})$. After kinetic decoupling ($s<1$),
$T_\chi/T_L\to\Gamma(\frac{3}{4})s^{1/4}\propto a^{-1}$.  Defining
the kinetic decoupling time by $s=1$ yields
\begin{eqnarray}\label{kindec}
  T_d&\equiv&T_L(s=1)=1.430\,C^{-1/4}g_{\rm eff}^{1/8}\left(\frac
    {m_\chi^5}{m_{\rm Pl}}\right)^{1/4}\nonumber\\
  &=&7.650\,C^{-1/4}g_{\rm eff}^{1/8}\left(\frac{m_\chi}{\hbox{100
    GeV}}\right)^{5/4}\ \hbox{MeV}\ .
\end{eqnarray}
For typical supersymmetry masses, kinetic decoupling occurs after
muon annihilation when the only abundant leptons are electron pairs
and neutrinos, for which (with $\sin^2\theta_W=0.223$)
$\sum_L(b_L^4+c_L^4)=\frac{13}{4}\tan^4\theta_W=0.268$ and $g_{\rm
eff}=43/4$. With $m_\chi=100$ GeV and $m_{\tilde L}=200$ GeV,
$C=0.0433$ yielding $T_d=22.6$ MeV.  Profumo {\it et al.}\
\cite{psk06} show that $T_d$ can span the range from a few MeV to a
few GeV for reasonable WIMP models. Here we take the supersymmetric
bino as a candidate but will show how numerical results scale with
$T_d$.

Next we examine the effect of density, velocity, and gravitational
potential fluctuations during kinetic decoupling.  The perturbed
phase space density is $f({\bf x},{\bf q},\tau)$, where ${\bf x}$
are comoving spatial coordinates, ${\bf q}=a{\bf p}$ are the
conjugate momenta, and $\tau$ is conformal time. The perturbed
line element in conformal Newtonian gauge is written
$ds^2=a^2[-(1+2\Phi)d\tau^2+(1-2\Psi)d{\bf x}\cdot d{\bf x}]$.
Including the effects of the metric perturbations $\Phi$ and
$\Psi$, the Boltzmann equation becomes
\begin{equation}\label{boltzpert}
  \frac{\partial f}{\partial\tau}+{\bf v}\cdot\frac{\partial f}
    {\partial{\bf x}}+\left(\dot\Psi{\bf q}-\epsilon\frac{\partial
    \Phi}{\partial{\bf x}}\right)\cdot\frac{\partial f}
    {\partial{\bf q}}=a\left(\frac{df}{dt}\right)_c\ ,
\end{equation}
where ${\bf v}={\bf q}/\epsilon$ is the proper velocity measured
by a comoving observer, $\epsilon\equiv(q^2+a^2 m_\chi^2)^{1/2}$
is the comoving energy, and ${\bf q}=q{\bf n}$.  With comoving
variables, the Fokker-Planck operator becomes
\begin{equation}\label{comboltzop}
  a\left(\frac{df}{dt}\right)_c=\gamma a\frac{\partial}
  {\partial{\bf q}}\cdot\left[({\bf q}-{\bf q}_L)f
  +a^2m_\chi T_L\frac{\partial f}{\partial{\bf q}}\right]\ ,
\end{equation}
where ${\bf q}_L\equiv am_\chi{\bf v}_L\equiv-am_\chi\nabla u_L$.

Now we linearize (\ref{comboltzop}) for first-order perturbations
of the lepton fluid by writing $T_L\to T_L(\tau)+T_1({\bf
x},\tau)$. The fields $T_1({\bf x},\tau)$, $\Phi({\bf x},\tau)$,
$\Psi({\bf x},\tau)$, and the lepton velocity potential $u_L({\bf
x}, \tau)$ are treated as first-order quantities. Assuming
$v^2\ll1$ and performing a spatial Fourier transform, we obtain
the linear perturbation equation
\begin{eqnarray}\label{cbeq1}
  \frac{1}{f_0}\left[\left(\frac{\partial f}{\partial\tau}
    \right)_q+i({\bf k}\cdot{\bf v})f-\gamma aL_{\rm FP}f
    \right]
    \ \ \ \ \ \ \ \ \ \ \ \ \ \ \ \ \ \ \ \ \ \ \\
  =\dot\Psi\frac{q^2}{\sigma_q^2}-\frac{i{\bf k}\cdot{\bf q}}
    {aT_\chi}(\Phi+\gamma au_L)+\gamma a\left(\frac{q^2}
    {\sigma_q^2}-3\right)\frac{T_1}{T_\chi}\ ,\nonumber
\end{eqnarray}
where
\begin{equation}\label{LFP}
  L_{\rm FP}f\equiv\frac{\partial}{\partial{\bf q}}\cdot
    \left({\bf q}f+a^2m_\chi T_L\frac{\partial f}{\partial{\bf q}}
    \right)\ .
\end{equation}
After WIMP freeze-out, the leptons dominate the gravitational
potentials so that $\dot\Psi$, $\Phi$, $u_L$, and $T_1$ are
functions of $({\bf k},\tau)$ given by the solution for a perfect
relativistic fluid.  Equation (\ref{cbeq1}) generalizes the
collisionless Boltzmann equation of Ref.\ \cite{mb95} to include the
effects of dark matter collisions with relativistic leptons.

To integrate (\ref{cbeq1}) for the phase space density $f$ we will
expand the momentum dependence in eigenfunctions of the
Fokker-Planck operator $L_{\rm FP}$:
\begin{subequations}\label{FPeigen}
\begin{eqnarray}
  L_{\rm FP}\phi_{nlm}&=&-(2n+l)\phi_{nlm}\ ,\label{FPeval}\\
  \phi_{nlm}&=&e^{-y}y^{l/2}L_n^{(l+1/2)}(y)Y_{lm}({\bf n})\ .
    \label{FPevec}
\end{eqnarray}
\end{subequations}
Here $y\equiv q^2/(2a^2m_\chi T_L)$ and $L_n^{(\alpha)}(y)$ is a
generalized Laguerre polynomial, also known as a Sonine
polynomial.  It is defined by the following series expansion:
\begin{equation}\label{sonexp}
  L_n^{(\alpha)}(x)=\sum_{k=0}^n\frac{\Gamma(n+\alpha+1)}
    {\Gamma(k+\alpha+1)}\frac{(-x)^k}{k!(n-k)!}\ .
\end{equation}
Generalized Laguerre polynomials have orthonormality relation
\begin{equation}\label{sonortho}
  \int_0^\infty\frac{n!\,x^\alpha e^{-x}}{\Gamma(n+\alpha+1)}
    L_n^{(\alpha)}(x)L_{n'}^{(\alpha)}(x)\,dx=\delta_{nn'}
\end{equation}
and completeness relation
\begin{equation}\label{soncompl}
  \sum_{n=0}^\infty\frac{n!\,(xx')^{\alpha/2}e^{-(x+x')/2}}
    {\Gamma(n+\alpha+1)}L_n^{(\alpha)}(x)L_n^{(\alpha)}(x')
    =\delta(x-x')\ .
\end{equation}

We will expand the phase space density $f({\bf k},{\bf q},\tau)$
in the complete set $\phi_{nlm}$.  However, it is unnecessary to
include all $(nlm)$ in this expansion.  Prior to kinetic
decoupling the Fokker-Planck operator $L_{\rm FP}$ rapidly damps
all terms except $\phi_{000}$ and those terms that are induced by
the right-hand side of (\ref{cbeq1}). The rotational symmetry of
this equation implies that only the $m=0$ terms are induced, where
the polar axis for the spherical harmonics is given by $\hat{\bf
k}={\bf k}/k$ \footnote{Collisional damping before neutrino
decoupling similarly justifies the neglect of the $m\ne0$ modes
for neutrinos in Ref.\ \cite{mb95}.}. Thus we may write
\begin{eqnarray}\label{laguerre}
  f({\bf k},{\bf q},\tau)
  &=&\frac{e^{-y}}{(2\pi a^2m_\chi T_L)^{3/2}}\nonumber\\
  &&\times\sum_{n,l=0}^\infty(-i)^l(2l+1)S_{nl}(y)P_l(\hat{\bf k}
    \cdot{\bf n})f_{nl}({\bf k},\tau)\ ,\nonumber\\
  S_{nl}&\equiv&y^{l/2}L_n^{(l+1/2)}(y)\ .
\end{eqnarray}

Before writing the perturbation equations, let us examine the
unperturbed solution, $f=f_0(q,\tau)\delta^3({\bf k})$.  The
exponential factors differ in (\ref{maxboltz}) and
(\ref{laguerre}), implying that the Laguerre expansion includes
more than one term.  Indeed, one finds
\begin{equation}\label{unpert}
  f=f_0(q,\tau)\quad\Rightarrow\quad f_{nl}=\delta_{l0}\left(
    1-\frac{T_\chi}{T_L}\right)^n\ .
\end{equation}
Prior to kinetic decoupling, when collisions maintain
$T_\chi=T_L$, $f_{00}=1$ and the other coefficients vanish. After
kinetic decoupling, $f_{n0}\to1$ for all $n$.  The Laguerre
expansion must be carried to high order in order to convert
$e^{-y}$ to $\exp(-q^2/2\sigma_q^2)$.  Similarly, we should expect
the perturbed phase space density also to require many terms in
$n$ after kinetic decoupling.


Substituting (\ref{laguerre}) into (\ref{cbeq1}) and using
orthonormality and several recurrence relations for the
generalized Laguerre polynomials, we obtain a system of coupled
ordinary differential equations for the evolution of the perturbed
phase space density,
\begin{eqnarray}\label{fnleom}
  \dot f_{nl}+(2n+l)(\gamma a+R)f_{nl}-2nR f_{n-1,l}
    \ \ \ \ \ \ \ \ \ \ \ \ \ \ \ \ \ \ \ \ \nonumber\\
  +k\sqrt{\frac{2T_L}{m_\chi}}\Bigl\{\textstyle\left(\frac{l+1}
    {2l+1}\right)\left[(n+l+\frac{3}{2})f_{n,l+1}
    -nf_{n-1,l+1}\right]\nonumber\\
  +\textstyle\frac{l}{2l+1}(f_{n+1,l-1}-f_{n,l-1})\Bigr\}
    \ \ \ \ \ \ \ \ \ \ \ \ \ \ \ \ \ \nonumber\\
  =\delta_{l0}[3\dot\Psi A_n-2(\dot\Psi+\gamma aT_1/T_\chi)
    B_n]
    \ \ \ \ \ \ \ \ \ \ \ \ \ \ \ \ \ \ \nonumber\\
  +\delta_{l1}\frac{k}{3}\sqrt{\frac{2m_\chi}{T_0}}\,
    (\Phi+\gamma au_L)A_n\ ,
    \ \ \ \ \ \ \ \ \ \ \ \ \ \ \ \ \
\end{eqnarray}
where
\begin{eqnarray}\label{ABRdef}
  R&\equiv&\frac{d}{d\tau}\ln(aT_L^{1/2})\ ,\nonumber\\
  A_n(\tau)&\equiv&\left(1-\frac{T_\chi}{T_L}\right)^n\ ,\nonumber\\
  B_n(\tau)&\equiv&n\left(\frac{T_\chi}{T_L}\right)
    \left(1-\frac{T_\chi}{T_L}\right)^{n-1}\ .
\end{eqnarray}
The density perturbation is $\delta({\bf k},\tau)=f_{00}$ for
${\bf k}\ne0$.

The source terms for Eqs.\ (\ref{fnleom}) are provided by the
relativistic plasma.  Their time-dependence changes during lepton
pair annihilation and neutrino decoupling.  For reasonable
parameters, neutralino kinetic decoupling occurs after muon
annihilation but before electron annihilation and neutrino
decoupling. During this era, the isentropic mode of perturbations
has time-dependence given by the relativistic perfect-fluid
transfer functions
\begin{subequations}\label{nucoupled}
\begin{eqnarray}
  \Phi=\Psi&=&-\frac{3}{\theta^3}(\sin\theta-\theta\cos\theta)\ ,\\
  {\cal H}u_L&=&-\frac{3}{2\theta}\left[\left(1-\frac{2}{\theta^2}\right)
    \sin\theta+\frac{2}{\theta}\cos\theta\right]\ ,\ \ \ \ \\
  \frac{T_1}{T_L}&=&\frac{\delta_L}{3}=-\frac{\theta^2}{2}\Phi
    -{\cal H}u_L\ ,
\end{eqnarray}
\end{subequations}
where ${\cal H}\equiv\dot a/a$, $\theta\equiv k\tau/\sqrt{3}$, and
the transfer functions are normalized so that $\Phi=-1$ for
$\theta=0$ \footnote{Here, $\delta_L=\delta\rho_L/(\rho_L+p_L)$ is
the number density perturbation in conformal Newtonian gauge; the
energy density perturbation for the relativistic leptons is
$\delta\rho_L/\rho_L=\frac{4}{3}\delta_L$.}. The actual
perturbations are obtained by multiplying the transfer functions
with the scalar field $\Psi_0({\bf k})$ that gives the spatial
dependence of the initial (inflationary) curvature fluctuations. As
the inflationary curvature perturbations are well known (a Gaussian
random field with nearly scale-invariant spectral density $P\propto
k^{-3}$), here we work with transfer functions.

Initial conditions for Eqs.\ (\ref{fnleom}) are obtained by
examining the solutions for $s=\frac{1}{2}\gamma a\tau\equiv
(\tau/\tau_d)^{-4}\gg 1$. Isentropic initial conditions have
$f_{00}=\delta=\delta_L$. The only significantly perturbed
components of $f$ for the strongly coupled plasma are the thermal
equilibrium values
\begin{equation}\label{initialcon}
  f_{00}=\delta_L\ ,\ \ f_{01}=\frac{k}{3}\sqrt{\frac{2m_\chi}
    {T_L}}\,u_L\ ,\ \ f_{10}=-\frac{T_1}{T_L}=-\frac{1}{3}\delta_L\ .
\end{equation}
All other components $f_{nl}$ are kept small by rapid WIMP-lepton
collisions.  Equations (\ref{fnleom}) with $n\le1400$ and $l\le15$
were integrated to $\tau=72\tau_d$ using the standard explicit
ordinary differential equation solver DVERK.  Convergence testing
showed that higher-order terms in the Laguerre expansion were
negligible.

\begin{figure}[t]
  \begin{center}
    \includegraphics[scale=0.7]{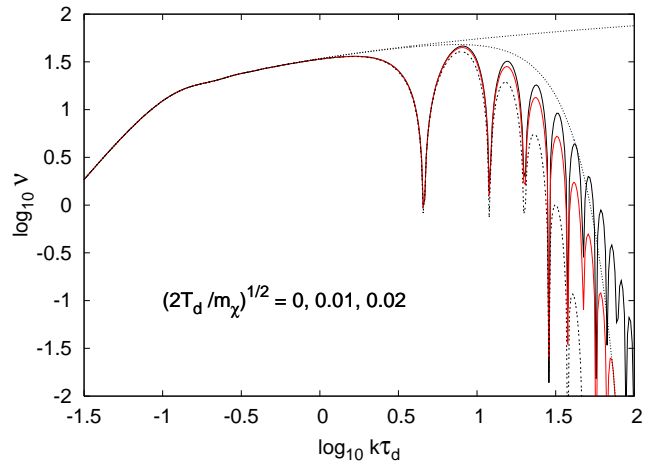}
  \end{center}
  \caption{CDM density transfer function versus wavenumber at conformal
    time $\tau=72\tau_d$.  The three oscillating curves assume that kinetic
    decoupling occurred at $\tau_d$ for three different values of the
    radiation temperature $T_d$ relative to the CDM particle mass
    $m_\chi$; the amplitude of the oscillations decreases with
    increasing $T_d/m_\chi$.  The upper, monotonic curve assumes that
    the CDM is always collisionless and was never coupled to the radiation.
    The other non-oscillating curve shows a crude model of kinetic
    decoupling described by a Gaussian cutoff.}
    \label{fig:fig1}
\end{figure}

Figure \ref{fig:fig1} shows the results for the density transfer
function expressed using the gauge-invariant variable $\nu$,
defined as the CDM number density perturbation in a synchronous
gauge for which the mean CDM velocity vanishes in the coordinate
frame (for a nonrelativistic particle, $\nu$ equals Bardeen's
variable $\epsilon_m$). This variable, which is used by CMBFAST
\cite{cmbfast}, is related to the conformal Newtonian gauge
variables by
\begin{equation}\label{epsilondef}
  \nu\equiv\delta+3{\cal H}u\ .
\end{equation}
For wavelengths longer than the radiation acoustic length
$\tau/\sqrt{3}$, $\nu=\delta_L+3{\cal H}u_L\approx
\frac{1}{2}(k\tau)^2$. For wavelengths shorter than the radiation
acoustic length at $\tau$ but longer than the acoustic length at
$\tau_d$ (i.e., $10\tau^{-1}<k<\tau_d^{-1}$), the acoustic
oscillations of the gravitational potential average out leading to a
suppression of growth induced in the CDM.  For these intermediate
wavelengths the transfer function is a logarithmic function of
wavenumber.  If the dark matter were completely non-interacting,
this logarithm would continue to arbitrarily high wavenumbers, as
illustrated by the monotonically rising curve in Fig.\
\ref{fig:fig1}.

Because WIMP dark matter was collisionally coupled to the
relativistic lepton plasma at early times, the CDM transfer function
in Fig.\ \ref{fig:fig1} shows remnant damped acoustic oscillations
at short wavelengths.  For comparison a Gaussian transfer function
is shown, with no oscillations. In this model, the effects of
kinetic decoupling are described by multiplying the transfer
function for the completely non-interacting case by
$\exp(-k^2/2\sigma_k^2)$.  As we will see, a simple model of free
streaming predicts that
$(\sigma_k\tau_d)^{-1}\approx[\Gamma(\frac{3}{4})T_d/m_\chi]^{1/2}
[\ln(\tau/\tau_d)]$ during the radiation-dominated era.  For
$\tau=72\tau_d$ and $(2T_d/m_\chi)^{1/2}=0.01$ this model would
predict $\sigma_k \tau_d\approx30$ whereas the curve shown in Fig.\
\ref{fig:fig1} has $\sigma_k\tau_d=18$.  Free-streaming does not
give a good approximation to the actual transfer function.

The exact transfer functions shown in Fig.\ \ref{fig:fig1}
decrease less rapidly with wavenumber than the approximate
transfer functions of Ref.\ \cite{lz05} which were computed using
a fluid approximation followed by free-streaming.  The following
section reviews the free-streaming solution and then develops a
more accurate approximation based on moments of the exact
Fokker-Planck equation.

\section{Approximate descriptions of perturbation evolution through
kinetic decoupling}
\label{sec:fluid}

In this section we consider two different approximations which
provide analytical insight to the numerical solution of the
Fokker-Planck equation.  The CDM behaves at early times like a
fluid and at late times like a free-streaming collisionless gas,
and in these limits we can develop useful analytical
approximations.

\subsection{Free-streaming model}

For $\tau\gg\tau_d$, the terms proportional to $\gamma a$ may be
dropped in (\ref{cbeq1}).  The differential equation may then be
integrated to give $f({\bf k},{\bf q},\tau)$ in terms of the initial
value $f({\bf k},{\bf q},\tau_\ast)$ for any $\tau_\ast\gg\tau_d$.
Integrating over momenta gives the conformal Newtonian gauge density
perturbation,
\begin{eqnarray}\label{freestream}
  \delta({\bf k},\tau)=\int d^3q\,e^{-i\zeta{\bf k}\cdot{\bf q}}\,
    f({\bf k},{\bf q},\tau_\ast)+\int_{\tau_\ast}^\tau d\tau'\,
    e^{-M'/2}\ \ \ \ \nonumber\\
  \times\left[(3-M')\dot\Psi({\bf k},\tau')+k^2\tau'
    \ln\left(\frac{\tau}{\tau'}\right)\Phi({\bf k},\tau')\right]
    \ ,\ \ \ \
\end{eqnarray}
where we have assumed evolution in the radiation-dominated era
with
\begin{equation}\label{zetadef}
  \zeta\equiv
    \frac{\tau_\ast}{a_\ast m_\chi}\ln\left(\frac{\tau}
    {\tau_\ast}\right)\ ,\ \ M'\equiv\frac{T_\chi'}{m_\chi}
    (k\tau')^2\ln^2\left(\frac{\tau}{\tau'}\right)\ .
\end{equation}
For large spatial frequencies, $k\tau\gg1$, the gravitational
potentials --- which are dominated by relativistic particles
--- oscillate rapidly leading to a small net integral
contribution; ignoring this, $\delta({\bf k},\tau)$ is a
momentum-space Fourier transform of the distribution function at
the initial time $\tau_\ast$.

Obtaining the exact solution still requires numerical integration of
(\ref{cbeq1}) through kinetic decoupling, or equivalently,
integrating the system of equations (\ref{fnleom}).  However, we can
get an idea of the effects of free-streaming by making an
instantaneous decoupling approximation, treating the CDM as a fluid
with a Maxwell-Boltzmann velocity distribution for $\tau<\tau_\ast$
and by (\ref{freestream}) for $\tau>\tau_\ast$, as was done in Ref.\
\cite{lz05}. Then, whether or not the CDM is strongly coupled to the
radiation, the perturbed distribution function is fully specified by
the perturbations of density, temperature, and velocity ${\bf
v}=-i{\bf k}u$,
\begin{equation}
  \delta f=f_0(q,\tau)\left[\frac{\delta\rho}{\rho}+\frac{1}{2}
    \left(\frac{q^2}{\sigma_q^2}-3\right)\frac{\delta T_\chi}
    {T_\chi}-\frac{iu{\bf k}\cdot{\bf q}}{aT_\chi}\right]\ .
\end{equation}
Carrying out the Fourier transform in (\ref{freestream}) now
yields
\begin{equation}\label{deltalz05}
  \delta(\tau)=e^{-M/2}\left[\left(\delta-\frac{M}{2}
    \frac{\delta T_\chi}{T_\chi}\right)_{\tau_\ast}
    -k^2\tau_\ast u(\tau_\ast)\ln
    \left(\frac{\tau}{\tau_\ast}\right)\right]\ ,
\end{equation}
where wavenumber arguments have been dropped for brevity, and
\begin{equation}\label{Mdef}
  M\equiv\frac{k^2}{k_f^2}\ln^2\left(\frac{\tau}{\tau_\ast}
    \right)\ ,\ \ k_f^{-2}=\frac{\tau_\ast^2 T_\chi(\tau_\ast)}
    {m_\chi}=\Gamma({\textstyle\frac{3}{4}})\frac{\tau_d^2T_d}
    {m_\chi}\ .
\end{equation}
If $\tau_\ast\sim\tau_d$, then $\delta T_\chi/T_\chi\approx
\frac{1}{3}\delta$; if $\tau_\ast\gg\tau_d$ and the fluid is
approximated as being adiabatic, $\delta T_\chi/T_\chi=\frac{2}{3}
\delta$.

Equation (\ref{deltalz05}) corrects errors in the definition of
$k_f$ of Ref.\ \cite{lz05} and adds the term proportional to $\delta
T_\chi/T_\chi$ to their Eq.\ (20). This new term arises from the
temperature perturbation of the CDM fluid, which modifies the
distribution function and, through free-streaming, modifies the
density for $\tau>\tau_\ast$. In particular, if the CDM
perturbations are approximately adiabatic, the temperature
perturbation causes the transfer function to decrease more rapidly
with $k$.

The free-streaming solution predicts a Gaussian cutoff of the
transfer function, $\delta\propto\exp(-k^2/2\sigma_k^2)$, with
cutoff distance equal to the free-streaming distance,
\begin{equation}\label{sigk}
  \sigma_k^{-1}(\tau)=\int_{\tau_\ast}^\tau\sqrt{\frac{T_\chi}
    {m_\chi}}\,d\tau=\left[\Gamma({\textstyle\frac{3}{4}})
    \frac{T_d}{m_\chi}\right]^{1/2}\int_{\tau_\ast}^\tau\frac{d\tau}
    {a/a_d}\ .
\end{equation}
During the radiation-dominated era, when $a\propto\tau$, the
free-streaming length grows logarithmically, but it saturates in the
matter-dominated era when $a\propto\tau^2$.  At first glance, Fig.\
\ref{fig:fig1} appears to qualitatively support the free-streaming
model of a Gaussian cutoff. However, the free-streaming model
predicts no damping for a super-heavy particle with
$T_d/m_\chi\to0$, while Fig.\ \ref{fig:fig1} shows that even in this
case there is damping.  This damping arises from friction between
the lepton and CDM gases during decoupling (Silk damping).  This
friction can be accounted for by treating the CDM as an imperfect
fluid.

\subsection{Imperfect fluid model}

To better describe an extended period of decoupling while allowing
for small deviations from a Maxwell-Boltzmann distribution, we
consider the evolution of the lowest order moments of the
distribution function.  We work to first order in perturbed
quantities and normalize the unperturbed distribution function to
$\int f_0\,d^3q=1$.  The perturbations for ${\bf k}\ne0$ then
define the lowest-order moments
\begin{eqnarray}\label{moments}
  \int f\,d^3q&=&\delta\ ,\nonumber\\
  \int fv_j\,d^3q&=&-ik_ju\ ,\nonumber\\
  \int fv_iv_j\,d^3q&=&(c_\chi^2\delta+\sigma)\delta_{ij}
    -\textstyle{\frac{3}{2}}
    (k_ik_j-\textstyle{\frac{1}{3}}k^2\delta_{ij})\pi \ ,\ \ \ \
\end{eqnarray}
where the density perturbation $\delta$, velocity potential $u$,
shear stress potential $\pi$, and entropy perturbation $\sigma$
are related to our expansion coefficients by
\begin{eqnarray}\label{fluidvars}
  &&\delta=f_{00}\ ,\ \ ku=\sqrt{\frac{9T_L}{2m_\chi}}\,f_{01}\ ,\ \
    k^2\pi=\frac{5T_L}{m_\chi}f_{02}\ ,\nonumber\\
  &&\ \ \ \ \ \ \ \ \sigma=\left(\frac{T_L}{m_\chi}-c_\chi^2
    \right)f_{00}-\frac{T_L}{m_\chi}f_{10}\ .
\end{eqnarray}
The effective sound speed squared of the CDM fluid is
\begin{equation}\label{cchi2}
  c_\chi^2=\frac{T_\chi}{m_\chi}\left(1-\frac{1}{3}\frac{d\ln
    T_\chi}{d\ln a}\right)\ ,
\end{equation}
which differs from the thermal speed squared $T_L/m_\chi$ appearing
in (\ref{fnleom}). This difference arises because the Laguerre
expansion uses eigenfunctions of the Fokker-Planck operator which
depends on the relativistic lepton temperature $T_L$ rather than the
WIMP temperature $T_\chi$. After kinetic decoupling, $T_\chi/T_L$
drops and the higher-order expansion coefficients $f_{nl}$ will
increase to compensate for this difference, as we already found
happening with the unperturbed distribution function in
(\ref{unpert}).

The variables in (\ref{fluidvars}) describe fluctuations of an
imperfect fluid. The reader may wonder how a single component fluid
can have an entropy perturbation.  A weakly imperfect fluid is
described by an equation of state $p=p(\rho,S)$ where $S$ is the
entropy which may vary in space and time.  However, the WIMP gas is
more complicated because it becomes fully collisionless after
kinetic decoupling; it may be regarded as a superposition of many
non-interacting ideal gases.

The time evolution of the imperfect fluid variables follows from
Eqs.\ (\ref{fnleom}) \footnote{These fluid equations, like the
Fokker-Planck equation from which they were derived, are correct
only to leading order in $T_L/m_\chi$.  The corrections to
(\ref{deldotf}) and (\ref{udotf}) are $\dot\Psi\to \dot\Psi-{\cal
H}\sigma$ and $\delta\to\nu=\delta+3{\cal H}u$, respectively.}:
\begin{subequations}\label{fluideqs}
\begin{eqnarray}
  &&\dot\delta+k^2u=3\dot\Psi\ ,\label{deldotf}\\
  &&\dot u+{\cal H}u=\Phi+c_\chi^2\delta+\sigma-k^2\pi-\gamma a
    (u-u_L)\ ,\ \ \ \ \label{udotf}\\
  &&\dot\sigma+2{\cal H}\sigma+\left(\frac{5}{3}\frac{T_L}{m_\chi}
    -c_\chi^2\right)k^2u-\frac{5}{4}\left(\frac{2T_L}{m_\chi}
    \right)^{3/2}kf_{11}\nonumber\\
  &&\ \ \ \ \ =-\frac{1}{a^2}\frac{d}{d\tau}(a^2c_\chi^2)\delta
    +3\dot\Psi\left(\frac{5}{3}\frac{T_\chi}{m_\chi}
    -c_\chi^2\right)\nonumber\\
  &&\ \ \ \ \ \ \ \ \ \ \ \ \ \  -2\gamma a\left[\sigma-\frac
    {T_1}{m_\chi}-\left(\frac{T_L}{m_\chi}-c_\chi^2\right)\delta
    \right]\ ,\label{sigdotf}\\
  &&\dot\pi+2{\cal H}\pi=\frac{4}{3}\frac{T_L}{m_\chi}u-\left(\frac
    {2T_L}{m_\chi}\right)^{3/2}\left(\frac{21}{4}
    \frac{f_{03}}{k}+\frac{f_{11}}{k}\right)\ \ \ \ \nonumber\\
  &&\ \ \ \ \ \ \ \ \ \ \ \ \ \ -2\gamma a\pi\ .\label{pidotf}
\end{eqnarray}
\end{subequations}
These equations are similar to those of an imperfect gas coupled to
the lepton fluid.  However, they differ significantly from the
Navier-Stokes equations assumed in Ref.\ \cite{hss01}.  In place of
a bulk viscosity term $\zeta k^2u$ and a shear viscosity term
$\frac{4}{3}\eta k^2u$ where $\zeta$ and $\eta$ are the bulk and
shear viscosity coefficients (divided by the mass density),
(\ref{udotf}) has an entropy term $\sigma$ and a shear stress term
$-k^2\pi$ where $\sigma$ and $\pi$ are not proportional to $u$. The
usual Chapman-Enskog expansion does not apply to our Fokker-Planck
equation when the collision mean-free path becomes large. Moreover,
because of the $f_{11}$ and $f_{03}$ terms, Eqs.\ (\ref{fluideqs})
do not form a closed system. Nonetheless, these equations are useful
for providing insight and they will guide us to a very good
approximation to the full numerical solution of the Fokker-Planck
equation.

Prior to kinetic decoupling, when the damping terms proportional to
$\gamma a$ are large, the solutions to (\ref{fluideqs}) have
$\delta=\delta_L$, $u=u_L$, $\sigma=\pi=0$, in agreement with
(\ref{initialcon}).  Entropy and shear stress perturbations develop
during decoupling as the CDM gas becomes collisionless. These in
turn modify and damp the acoustic oscillations of the gas.

It is instructive to solve the imperfect fluid equations with
several different approximations, in order to deduce which
physical effects are responsible for the features of the transfer
functions shown in Fig.\ \ref{fig:fig1}.  The most extreme
approximation is to completely neglect the CDM temperature and
coupling to other particles, so as to describe a perfect cold,
collisionless gas. This approximation consists of setting
$c_\chi^2=\sigma=\pi=\gamma=0$ retaining only the gravitational
interaction between the CDM and the relativistic plasma.  In this
case the exact solution prior to neutrino decoupling for
isentropic initial conditions is
\begin{eqnarray}\label{ccfsol}
  \frac{\nu}{9}&=&\frac{\cos\theta+\theta\sin\theta}{\theta^2}
    -\hbox{Ci}(\theta)+\ln\theta-\frac{1}{\theta^2}+{\cal C}-\frac{1}{2}
    \ ,\nonumber\\
  {\cal H}u&=&\frac{3}{\theta^3}(\sin\theta-\theta)\ ,
\end{eqnarray}
where $\theta=k\tau/\sqrt{3}$, ${\cal C}=0.5772\cdots$ is the
Euler-Mascharoni constant, and $\hbox{Ci}(\theta)$ is the cosine
integral.  This solution is shown by the monotonically increasing
curve in Fig.\ \ref{fig:fig1}. We see that once the wavelength
becomes smaller than the relativistic acoustic horizon and the
potentials oscillate faster than the CDM can respond, the CDM
density perturbation growth slows to logarithmic in time.  This
suppression is responsible for the turnover of the low-redshift
CDM power spectrum from $P(k)\propto k$ at long wavelengths to
$P(k)\propto k^{-3}\ln^2(k)$ at short wavelengths.  However, the
approximation of a cold, collisionless fluid includes none of the
physical effects of kinetic decoupling.

The next simplest approximation is to treat the CDM as being cold
($c_\chi^2=\sigma=\pi=0$) but to include the friction term
$-\gamma a (u-u_L)$ in the velocity equation.  This approximation
is exact in the limit $T_d/m_\chi\to0$ hence it reproduces the
case $T_d/m_\chi=0$ in Fig. \ref{fig:fig1}.  We can find the
solution in the limit $\tau\gg\tau_d$ by first noting that the
general solution of the second-order system $\dot\delta+k^2u=
3\dot\Psi$, $\dot u+{\cal H}u=\Phi$ is given by
\begin{eqnarray}\label{coldlimsol}
  \frac{\nu}{9}&=&\frac{\cos\theta+\theta\sin\theta}{\theta^2}
    -\hbox{Ci}(\theta)+f_1\left(\ln\theta-\frac{1}{\theta^2}\right)
    +f_2\ ,\nonumber\\
  {\cal H}u&=&\frac{3}{\theta^3}(\sin\theta-f_1\theta)\ ,
\end{eqnarray}
where $f_1$ and $f_2$ are independent of $\tau$ but may depend on
$k$.   Including the damping terms in (\ref{udotf}) promotes $f_1$
and $f_2$ to functions $f_1(k,\tau)$ and $f_2(k,\tau)$ obeying the
differential equations
\begin{subequations}\label{f12dot}
\begin{eqnarray}
  \dot f_1+\gamma a f_1&=&\gamma\left(\cos\theta+\frac{1}{2}
    \theta\sin\theta\right)\ ,\label{f1dot}\\
  \dot f_2+\dot f_1\ln\theta&=&0\ .\label{f2dot}
\end{eqnarray}
\end{subequations}
In the strongly-coupled limit $\tau\ll\tau_d$ the solution must
match Eqs.\ (\ref{nucoupled}), which gives
\begin{eqnarray}\label{earlycoldlimsol}
  f_1(k,\tau)&\to&\cos\theta+\frac{1}{2}\theta\sin\theta\ ,
    \nonumber\\
  f_2(k,\tau)&\to&\hbox{Ci}(\theta)-\left(\cos\theta+\frac{1}{2}
    \theta\sin\theta\right)\ln\theta\ \ \ \ \nonumber\\
    &&\ \ \ \ \ \ \ \ -\frac{1}{2}\cos\theta\ .
\end{eqnarray}
The exact solution of (\ref{f12dot}) satisfying these initial
conditions is given by a pair of quadratures,
\begin{eqnarray}\label{coldfric}
  f_1(k,\tau)&=&\int_0^\tau e^{(s-s')/2}\left(\cos\theta'
    +\frac{1}{2}\theta'\sin\theta'\right)\gamma' a'\,d\tau'
    \ ,\nonumber\\
  f_2(k,\tau)&=&\int_0^\tau[f_1(k,\tau')-1]\,\frac{d\tau'}{\tau'}
    +(1-f_1)\ln\theta\nonumber\\
  &&\ \ \ \ \ \ \ \ \ \ \ \ \ \ \ \ +{\cal C}-\frac{1}{2}\ ,
\end{eqnarray}
where $s\equiv\frac{1}{2}\gamma a\tau=(\tau/\tau_d)^{-4}$,
$\theta\equiv k\tau/\sqrt{3}$, and primed quantities are evaluated
at $\tau'$.

\begin{figure}[t]
  \begin{center}
    \includegraphics[scale=0.85]{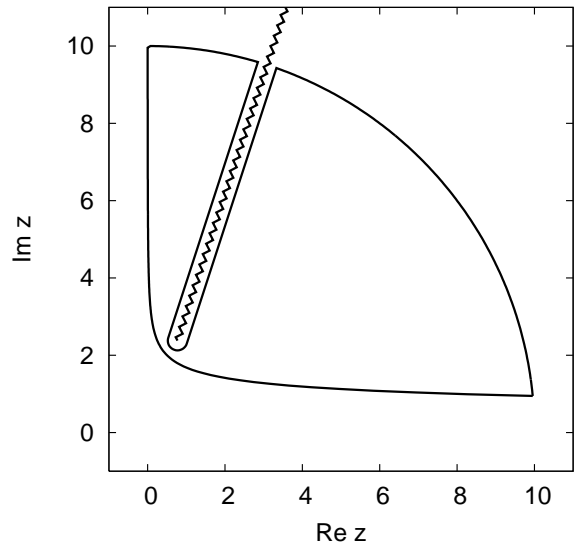}
  \end{center}
  \caption{Contour used to evaluate the integral in Eq.\
  (\ref{f1int}).  The quarter-circle is actually taken to have
  a much larger radius than shown so that its contributions
  to the contour integral vanish.  The desired path for the
  integral is along the lower left curve that joins the
  quarter-circle.}
  \label{fig:f1cont}
\end{figure}

\begin{table}[t]
\begin{tabular}{|rr|rr|rr|rr|rr|}
  \hline
  $x\ $ & $f_{\rm res}\quad$ & $x\ $ & $f_{\rm res}\quad$ & $x\ $ & $f_{\rm res}\quad$
                            & $x\ $ & $f_{\rm res}\quad$ & $x\ $ & $f_{\rm res}\quad$\\
  \hline
  0.0 &   1.0000 & 1.0 &$-$0.1637 & 2.0 & 0.0252 & 3.0 &   0.0099 & 4.0 &$-$0.0101\\
  0.1 &   0.5744 & 1.1 &$-$0.1562 & 2.1 & 0.0345 & 3.1 &   0.0039 & 4.1 &$-$0.0082\\
  0.2 &   0.3791 & 1.2 &$-$0.1409 & 2.2 & 0.0403 & 3.2 &$-$0.0013 & 4.2 &$-$0.0062\\
  0.3 &   0.2275 & 1.3 &$-$0.1202 & 2.3 & 0.0427 & 3.3 &$-$0.0056 & 4.3 &$-$0.0042\\
  0.4 &   0.1063 & 1.4 &$-$0.0965 & 2.4 & 0.0424 & 3.4 &$-$0.0090 & 4.4 &$-$0.0022\\
  0.5 &   0.0109 & 1.5 &$-$0.0716 & 2.5 & 0.0397 & 3.5 &$-$0.0113 & 4.5 &$-$0.0004\\
  0.6 &$-$0.0610 & 1.6 &$-$0.0471 & 2.6 & 0.0352 & 3.6 &$-$0.0126 & 4.6 &   0.0011\\
  0.7 &$-$0.1196 & 1.7 &$-$0.0244 & 2.7 & 0.0295 & 3.7 &$-$0.0130 & 4.7 &   0.0023\\
  0.8 &$-$0.1443 & 1.8 &$-$0.0045 & 2.8 & 0.0231 & 3.8 &$-$0.0126 & 4.8 &   0.0033\\
  0.9 &$-$0.1606 & 1.9 &   0.0122 & 2.9 & 0.0164 & 3.9 &$-$0.0116 & 4.9 &   0.0039\\
  \hline
\end{tabular}
  \caption{Residuals from the steepest descent approximation to
  $f_1(x)$, defined in Eqs.\ (\ref{f1int}) and (\ref{f1cont}).}
  \label{tab:tab1}
\end{table}

The late-time solution is dominated by $f_1$, which for
$\tau\gg\tau_d$ becomes the following function of wavenumber,
\begin{equation}\label{f1int}
  f_1(x)={\cal R}e\!\!\int\!\!e^{-z(r)}\frac{\left(1+xre^{-i2\pi/5}\right)}
    {r^5-1}\,dz\ ,\ \
    x\equiv\left(\frac{k\tau_d}{2\sqrt{3}}\right)^{4/5}\!.
\end{equation}
The complex function $z(r)=xe^{-i2\pi/5}(\frac{1}{2}r^{-4}+2r)$
where $r$ itself is complex but cannot be used as the integration
variable because of the essential singularity at $r=0$. The
contour used to evaluate $f_1$ is shown in Fig.\ \ref{fig:f1cont}.
Using the steepest descent approximation to evaluate the
contributions along the branch cut gives
\begin{eqnarray}\label{f1cont}
  f_1(x)&=&\left(\frac{4\pi x}{5}\right)^{1/2}\exp\left[-\frac{5}
    {2}x\cos\left(\frac{2\pi}{5}\right)\right]\nonumber\\
  &&\times\left[\cos\left(\varphi-\frac{\pi}{5}\right)+x\cos
    \left(\varphi-\frac{3\pi}{5}\right)\right]+f_{\rm res}(x)
    ,\nonumber\\
  \varphi&\equiv&\frac{5x}{2}\sin\left(\frac{2\pi}{5}\right)\ ,\ \
\end{eqnarray}
where $f_{\rm res}(x)$ is a correction to the steepest descent
approximation, evaluated numerically and tabulated for
interpolation in Table \ref{tab:tab1}. For $x\ll1$, $1-f_1$ goes
to zero faster than $x^4$, with $f_1(0.2)=0.9990$.

\begin{figure}[t]
  \begin{center}
    \includegraphics[scale=0.7]{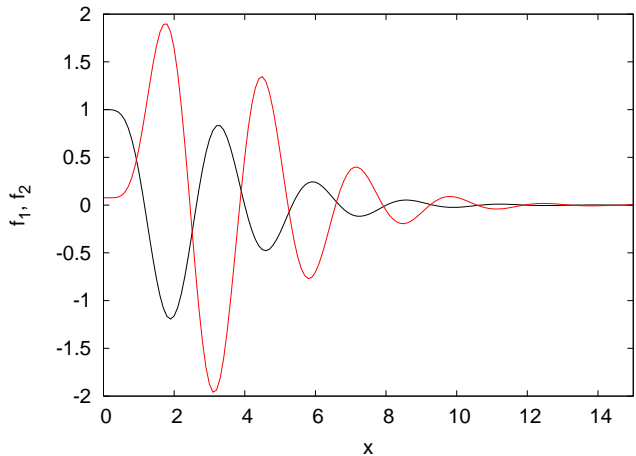}
  \end{center}
  \caption{The auxiliary functions $f_1(x)$ and $f_2(x)$ (the curve with
  the higher amplitude of oscillation) appearing in Eq.\ (\ref{coldlimsol}),
  where $\tau\gg\tau_d$ and $x\equiv(k\tau_d/2\sqrt{3})^{4/5}$.}
  \label{fig:fig3}
\end{figure}

Numerical integration yields an approximate fit to $f_2(x)$ for
$x\gg1$ similar to (\ref{f1cont}):
\begin{equation}\label{f2fit}
  f_2(x)\approx-3.45x^{0.75}\exp\left[-\frac{5}{2}x\cos\left(
    \frac{2\pi}{5}\right)\right](x\sin\varphi-\cos\varphi)\ ,
\end{equation}
Figure \ref{fig:fig3} shows the results for $f_1(x)$ and $f_2(x)$.
Although $f_2$ has larger amplitude, $f_1$ dominates the
contribution to the CDM transfer function for $\tau\gg\tau_d$.

This calculation shows that, in the limit of large WIMP mass
$m_\chi\gg T_d$, the free-streaming suppression of the transfer
function is not Gaussian but instead is exponential in $x\propto
k^{4/5}$.

Allowing a nonzero $T_d/m_\chi$ introduces a Gaussian suppression,
as follows.  Including the $c_\chi^2\delta$ term in (\ref{udotf})
adds the term $-k^2c_\chi^2\tau\nu/9$ to the right-hand side of
(\ref{f1dot}).  With the leading late-time behavior $\epsilon/9
\approx f_1\ln\theta$, this gives an approximate Gaussian
suppression,
\begin{equation}\label{tempcut}
  \nu(k,\tau)\approx\exp\left(-\frac{1}{2}\frac{k^2}{k_{s}^2}
    \right)\nu_0(k,\tau)\ ,\ \
    k_{s}^{-1}\equiv\int_0^\tau c_\chi(\tau')\,d\tau'\ ,
\end{equation}
where $\nu_0(k,\tau)$ is the solution for $T_d/m_\chi=0$ given by
(\ref{coldlimsol}) and (\ref{coldfric}).  While (\ref{tempcut}) is
similar to (\ref{deltalz05}) and (\ref{Mdef}), the Gaussian cutoff
distance differs from the free-streaming distance (\ref{sigk}) by a
factor $\sqrt{5/3}$.

\begin{figure}[t]
  \begin{center}
    \includegraphics[scale=0.7]{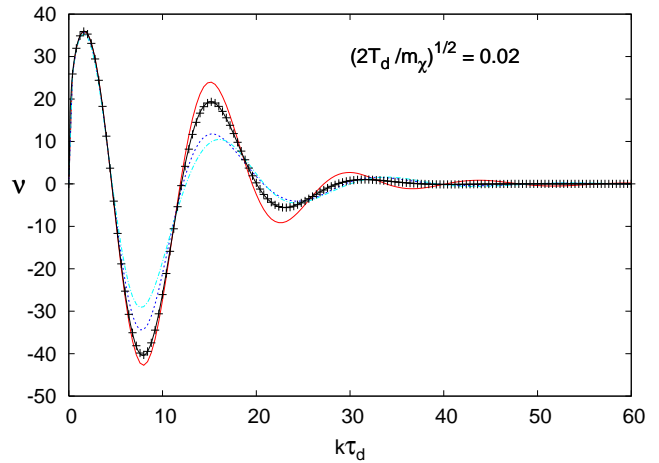}
  \end{center}
  \caption{CDM transfer function and several approximations
    plotted versus wavenumber at conformal time $\tau=72\tau_d$.
    In descending amplitude of the second peak, the curves are
    (1) fluid approximation with $\pi=\sigma=0$, (2) exact
    solution of the Fokker-Planck equation, (3) imperfect
    fluid approximation with shear stress but no entropy
    perturbation; (4) imperfect fluid approximation with both
    shear stress and entropy perturbations.  The plus signs
    superimposed on the exact solution curve are the solution
    with $T_d/m_\chi=0$, multiplied by a Gaussian damping factor
    as described in the text.}
  \label{fig:fig4}
\end{figure}

Equations (\ref{coldlimsol}) and (\ref{tempcut}) assume that the CDM
behaves like a perfect fluid \footnote{For example, Eq.\
(\ref{tempcut}) describes the small-scale damping of fluctuations
after recombination in a baryon-dominated universe.}.  By
integrating (\ref{fluideqs}) numerically with the simplification
$f_{11}=f_{03}=0$, we can include effects of nonzero shear stress
$\pi$ and entropy perturbation $\sigma$. The results are shown in
Fig.\ \ref{fig:fig4} in comparison with the exact solution from
Fig.\ \ref{fig:fig1}. While the effects of nonzero temperature,
shear stress, and entropy perturbations qualitatively reproduce the
suppression of the transfer function, none of the fluid
approximations gives a good match to the exact solution of the
Fokker-Planck equation. However, an excellent fit (with maximum
error about one percent of the oscillation amplitude) is given by a
modification of (\ref{tempcut}),
\begin{eqnarray}\label{tempfit}
  \nu(k,\tau)&\approx&\exp\left(-\frac{1}{2}\frac{k^2}{k_{\rm fs}^2}
    \right)\nu_0(k,\tau)\ ,\nonumber\\
    k_{\rm fs}^{-1}&\equiv&\sqrt{\frac{6T_d}{5m_\chi}}
      \int_{\tau_\ast}^\tau\frac{d\tau}{a/a_d}\ .
\end{eqnarray}
This approximation, with $\tau_\ast=1.05\tau_d$, is shown by the
plus signs in Fig.\ \ref{fig:fig4}.  The coefficient in front of the
integral defining $k_{\rm fs}^{-1}$ was found numerically; the value
$\frac{6}{5}=1.200$ is numerically correct to 0.1\% or better and
clearly differs from the coefficient
$\Gamma(\frac{3}{4})=1.2254\cdots$ appearing in the free-streaming
prediction of (\ref{sigk}) as well as the perfect fluid prediction
$\frac{5}{3}\Gamma(\frac{4}{3})=2.0423\cdots$ of (\ref{tempcut}).

The modified form of thermal damping suggests that on small scales
the WIMP gas might be described as a thermal gas with ratio of
specific heats $1.2/\Gamma(\frac{4}{3})=0.979\cdots$ instead of
$\frac{5}{3}$. However, numerical tests showed that no perfect gas
equation of state can reproduce the exact solution of the
Fokker-Planck equation as well as (\ref{tempfit}).  We will
therefore use (\ref{tempfit}) to calculate the effects of
free-streaming damping even though it is based on a numerical
instead of an analytic solution.  Note that (\ref{tempfit}) is valid
through pair annihilation and radiation-matter equality because the
free-streaming distance is proportional to
$\int\sqrt{T_\chi/m_\chi}\,d\tau$ and $T_\chi\propto(a/a_d)^{-2}$ at
all times after WIMP decoupling.

The physics of WIMP decoupling is similar but not identical to the
much later decoupling of atoms at a temperature of 0.25 eV. In both
cases the decoupled particles bear the imprint of acoustic
oscillations while they were coupled to a relativistic gas.  In both
cases the acoustic oscillations are damped by friction during
decoupling (Silk damping), although the damping is exponential in
$k^{4/5}$ for WIMPs as opposed to $k^2$ for atoms. In both cases
short-wavelength fluctuations are damped further by thermal motions
after decoupling. However, the last stage differs for the two gases
because the atomic gas remains collisional (hence damping takes
place for wavelengths shorter than the Jeans length) while the CDM
gas is collisionless and the relevant scale is the free-streaming
length. As a result the CDM gas develops shear stress and entropy
perturbations in (\ref{udotf}) that are not present for a
collisional gas.  Nonetheless these perturbations lead to
free-streaming damping that is qualitatively similar to the Jeans
damping of a collisional gas.

\begin{figure}[t]
  \begin{center}
    \includegraphics[scale=0.7]{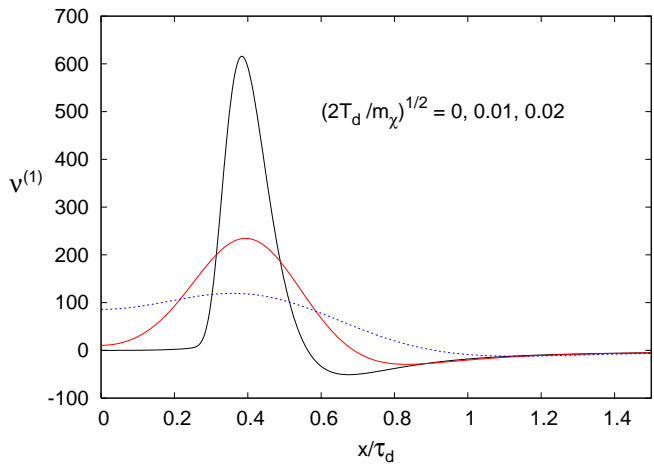}
  \end{center}
  \caption{Real space CDM Green's function (Fourier transform of
    the transfer function) at conformal time $\tau=10^7\tau_d$,
    approximately at the end of the radiation-dominated era.
    An initial planar perturbation sends an acoustic wave through
    the relativistic plasma.  This wave travels through the CDM
    until kinetic decoupling ends; thereafter the wave is frozen
    in place but grows logarithmically in amplitude.  The three
    curves show the effect of diffusion with increasing CDM
    temperature.}
  \label{fig:figgr1}
\end{figure}

The physics of perturbation evolution can be simpler to interpret
in position space than in Fourier space \cite{babert02}.  Figure
\ref{fig:figgr1} shows the one-dimensional Green's function
defined by
\begin{equation}\label{green1}
  \nu^{(1)}(x,\tau)=\frac{1}{2\pi}\int_{-\infty}^\infty
    e^{i{\bf k}\cdot{\bf x}}\,\nu(k,\tau)\,dk\ .
\end{equation}
In real space the Green's function is essentially a wave packet of
sound that started at $x=0$ and propagated until kinetic decoupling
after which time it froze in place.  The oscillations in $k$-space
arise solely because the wave packet in real space has a rapid
change in slope at its trailing edge. In the absence of WIMP-lepton
coupling, $\nu^{(1)}$ would have a delta function singularity at
$x=0$ and a compensating underdense tail at $x>0$ due to the
gravitational perturbation caused by the outgoing acoustic wave in
the relativistic plasma \cite{babert02}.

Until now we have assumed that the photon-lepton plasma is a perfect
relativistic gas with $p=\frac{1}{3}\rho$.  This approximation
breaks down after neutrino decoupling and during electron/positron
pair annihilation.  The effects of neutrino free-streaming are small
and are already included in numerical codes such as CMBFAST and
COSMICS but pair annihilation is not. As we show next, pair
annihilation also results in a minor modification of the CDM
transfer function on small scales.

\bigskip
\section{Evolution through Pair Annihilation}

During electron-positron annihilation the equation of state
changes slightly, modifying the evolution of perturbations.
Neglecting the small amount of neutrino heating that takes place
during pair annihilation, after CDM kinetic decoupling the energy
density and pressure of the photons, 3 flavors of neutrinos, and
electron pairs is
\begin{eqnarray}\label{gamnue}
  \rho_\gamma=3p_\gamma=\frac{\pi^2T_\gamma^4}{15}\ ,&&
  \rho_\nu=3p_\nu=\frac{7\pi^2T_{\nu 0}^4}{40a^4}\ ,
    \nonumber\\
  \rho_\pm=\frac{2T_\gamma^4}{\pi^2}R(\xi)\ ,&&
  p_\pm=\frac{2T_\gamma^4}{3\pi^2}P(\xi)\ ,
\end{eqnarray}
where $\xi\equiv m_e/T_\gamma$ and $T_{\nu 0}\equiv aT_\nu
=\hbox{constant}$.  The pair density and pressure are given in terms
of the Fermi-Dirac integrals
\begin{eqnarray}\label{RPdef}
  R(\xi)&\equiv&\int_0^\infty\frac{x^2\sqrt{x^2+\xi^2}\,dx}
    {e^{\sqrt{x^2+\xi^2}}+1}\ ,\nonumber\\
  P(\xi)&\equiv&\int_0^\infty\frac{x^4\,dx}{\sqrt{x^2+\xi^2}
    (e^{\sqrt{x^2+\xi^2}}+1)}\ \ .
\end{eqnarray}
Energy conservation for the photon-pair plasma gives
\begin{equation}\label{yprime}
  \frac{d\ln\xi}{d\ln a}=\frac{A}{B}\ ,
\end{equation}
where
\begin{eqnarray}\label{ABdef}
  A&\equiv&1+\frac{30R}{\pi^4}+\frac{15}{2\pi^4}(P-R)\ ,
    \nonumber\\
  B&\equiv&1+\frac{30R}{\pi^4}-\frac{15}{2\pi^4}
    \frac{dR}{d\ln y}\ .
\end{eqnarray}
The photon-pair plasma has equation of state parameter
\begin{equation}\label{wtot}
  w=\frac{1}{3}+\frac{(10/\pi^4)(P-R)}{1+(30/\pi^4)R
    +(21/8)(T_\nu/T_\gamma)^4}
\end{equation}
and sound speed squared
\begin{equation}\label{c2tot}
  3c_w^2=1+\frac{A(A-B)}{B[A+(21/8)(T_\nu/T_\gamma)^4]}\ .
\end{equation}
\begin{figure}[t]
  \begin{center}
    \includegraphics[scale=0.7]{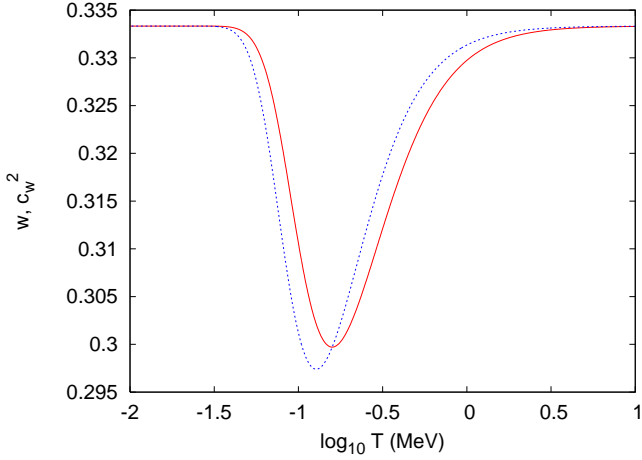}
  \end{center}
  \caption{Equation of state parameter $w$ (solid curve)
    and effective sound speed squared $c_w^2$ (dotted curve)
    through the period of electron pair annihilation.}
  \label{fig:figeos}
\end{figure}
These are plotted in Fig.\ \ref{fig:figeos}.  Pair annihilation
makes a 10\% dip in the equation of state.

\begin{figure}[t]
  \begin{center}
    \includegraphics[scale=0.65]{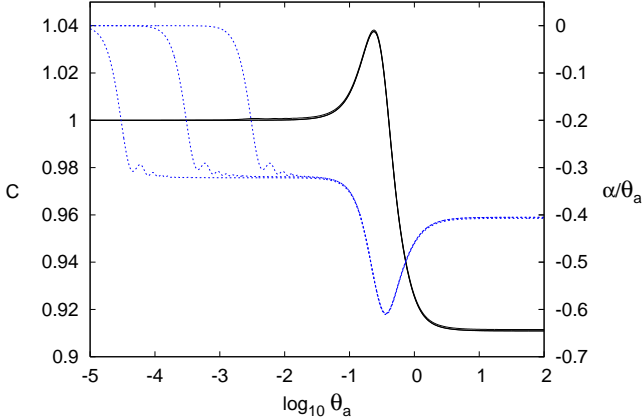}
  \end{center}
  \caption{Amplitude $C$ (solid curve) and phase shift $\alpha$
    (dashed curves, normalized by $\theta_a=k\tau_a/\sqrt{3}$)
    for the gravitational potential $\Psi$ of Eq.\ (\ref{Phipa}),
    for $\tau\gg\tau_a$, plotted versus $\theta_a$ (dimensionless
    wavenumber).  Three sets of curves are shown, corresponding
    to the times $\tau=(10^3,10^4,10^5)\,\tau_a$.  Before pair
    annihilation, $C=1$ and $\alpha=0$ for all $k$.}
  \label{fig:figCa}
\end{figure}

The relation between expansion factor and conformal time follows
from integrating the Friedmann equation
\begin{equation}\label{friedmann}
  \dot a^2=\frac{4\pi^3 G(aT_\gamma)^4}{45}g_{\rm eff}\ ,
\end{equation}
where
\begin{equation}\label{geff}
  g_{\rm eff}=\frac{30}{\pi^2}\left(\frac{\rho_\gamma
    +\rho_\nu+\rho_\pm}{T_\gamma^4}\right)
  =2+\frac{21}{4}\left(\frac{T_\nu}{T_\gamma}\right)^4
    +\frac{60}{\pi^4}R\ .
\end{equation}
Neglecting the small neutrino shear stress arising from
free-streaming after neutrino decoupling at $T\sim2$ MeV, the
conformal Newtonian gauge gravitational potentials $\Phi=\Psi$
obey the evolution equation
\begin{equation}\label{Phieom}
  \ddot\Phi+3(1+c_w^2){\cal H}\dot\Phi+3(c_w^2-w){\cal H}^2
    \Phi+k^2c_w^2\Phi=0\ .
\end{equation}
Long after pair annihilation is completed, $w=c_w^2=\frac{1}{3}$
and the solution is
\begin{equation}\label{Phipa}
  \Phi(k,\tau)=-\frac{3C}{\theta^3}[\sin(\theta+\alpha)-\theta
    \cos(\theta+\alpha)]\ ,
\end{equation}
where $C$ (not to be confused with the collisional coupling constant
of Eq. \ref{Cneut}) and $\alpha$ are independent of $\tau$ but may
depend on $k$. Before pair annihilation, $C=1$ and $\alpha=0$.
During pair annihilation, $C$ and $\alpha$ change but after pair
annihilation they become independent of time.

We define a characteristic conformal time for pair annihilation,
$\tau_a$, by scaling from the decoupling time:
\begin{equation}\label{taua}
  \tau_a\equiv\tau_d\left(\frac{T_d}{\hbox{0.511 MeV}}\right)\ .
\end{equation}
Pair annihilation imprints features on $\Phi(k,\tau)$ at wavenumbers
$k\sim\tau_a^{-1}$.  After pair annihilation is completed, we expect
$C$ and $\alpha$ in (\ref{Phipa}) to depend only on $k\tau_a$.
Numerical integration of (\ref{Phieom}) yields the results shown in
Fig.\ \ref{fig:figCa}. Pair annihilation leads to a small change in
the amplitude $C$ of gravo-acoustic oscillations and a
characteristic phase shift $\alpha$ of order $\theta_a\equiv
k\tau_a/\sqrt{3}$.  Wavelengths much longer than the acoustic
horizon length ($\theta=k\tau/\sqrt{3}\ll1$) are unmodified. Waves
that enter the Hubble length when the effective sound speed is
reduced by pair annihilation (Fig.\ \ref{fig:figeos}) are amplified
because with decreasing sound speed, pressure forces are less able
to prevent gravitational growth.  The diminished sound speed also
changes the distance traveled by acoustic waves, leading to a phase
shift for wavelengths smaller than the acoustic horizon distance.
The propagation of the acoustic horizon is evident in the
propagating steps in $\alpha/\theta_a$ in Fig.\ \ref{fig:figCa}.
Although these steps would appear to prevent $\Phi$ from relaxing to
(\ref{Phipa}) with time-independent $C$ and $\alpha$, for
$\tau\gg\tau_a$, $\theta\gg|\theta_a|$ and one may set $\alpha=0$
with an error in $\Phi$ of $O(\tau_a/\tau)$.  Wavelengths much
longer than the Hubble length during pair annihilation are
essentially unmodified.

Still shorter waves are described by the WKB solution of
(\ref{Phieom}),
\begin{equation}\label{wkbpsi}
  \Phi(k,\tau)\propto\left(\frac{\rho+p}{c_w}\right)^{1/2}
    \cos\int_0^\tau kc_w'\,d\tau'\ \ \hbox{for}\ \ k\tau\gg1\ .
\end{equation}
The dip in the sound speed during pair annihilation leads to a
phase shift $\alpha=k\int_0^\tau(c_w'-1/\sqrt{3})\,d\tau'
=-0.405\theta_a$ for $\theta_a\gg1$ and $\tau\gg\tau_a$. Pair
annihilation effectively resets the starting time for acoustic
oscillations from $\tau=0$ to $\tau=0.405\tau_a$.

Similarly the change in energy density and pressure through pair
annihilation leads to a change in amplitude.  In the WKB
approximation, one finds that for $k\tau_a\gg1$ the amplitude $C$ of
$\Phi(k,\tau)$ changes through pair annihilation by a factor
\begin{equation}\label{Cwkb}
  C(k)\to\frac{\dot a(\tau_1)}{\dot a(\tau_2)}=
  \left(\frac{T_\nu}{T_\gamma}\right)^2_{\!\!\tau_2}\left[
    \frac{g_{\rm eff}(\tau_1)}{g_{\rm eff}(\tau_2)}\right]^{1/2}
    =0.911\ ,
\end{equation}
where $\tau_1\ll\tau_a\ll\tau_2$. The WKB results for $\alpha(k)$
and $C(k)$ match the numerical results shown in Fig.\
\ref{fig:figCa}.

The change in the gravitational potential caused by pair
annihilation induces changes in the CDM growth.  The late-time
solution of (\ref{deldotf})--(\ref{udotf}) with $T/m_\chi=0$ is
\begin{eqnarray}\label{delupa}
  \frac{\nu_0}{9C}&=&\frac{\cos\tilde\theta+\theta\sin\tilde\theta}
    {\theta^2}-\cos\alpha\,\hbox{Ci}(\theta)
    +\sin\alpha\,\hbox{Si}(\theta)\nonumber\\
  &&\ \ \ \ \ \ \ \ \
    +f_3\left(\ln\theta-\frac{1}{\theta^2}\right)
    +f_4\ ,\nonumber\\
  {\cal H}u_0&=&\frac{3C}{\theta^3}[\sin\tilde\theta-f_3\theta]\ ,
\end{eqnarray}
where $\tilde\theta\equiv\theta+\alpha$ and the subscript 0
indicates $T/m_\chi=0$.  After pair annihilation the transfer
functions $f_3$ and $f_4$ depend on $k$ but not on $\tau$.

\begin{figure}[t]
  \begin{center}
    \includegraphics[scale=0.65]{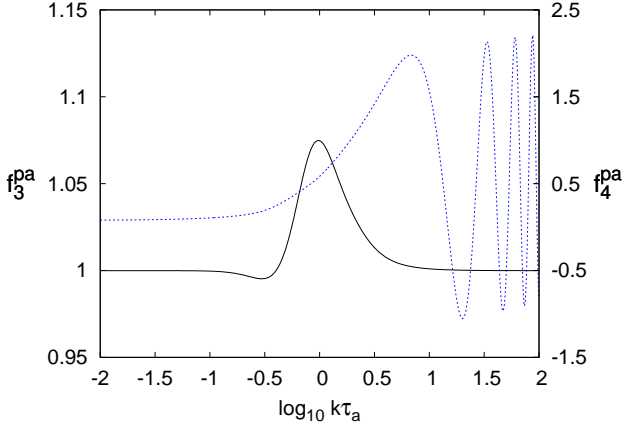}
  \end{center}
  \caption{Amplitude functions $f_3^{\rm pa}$ (solid curve) and
    $f_4^{\rm pa}$ (dashed curve) for the gravitational CDM density
    perturbation $\nu$ of Eq.\ (\ref{delupa}), for $\tau\gg\tau_a$,
    plotted versus wavenumber.  For $k\tau_a\ll1$, $f_3^{\rm pa}=1$
    and $f_4^{\rm pa}={\cal C}-\frac{1}{2}=0.0772\cdots$.  This figure
    assumes that the CDM is always collisionless and was never coupled
    to the radiation.  The results show there is a 7\% enhancement of
    growth for waves that enter the horizon during pair annihilation.}
  \label{fig:figfgam0}
\end{figure}

Kinetic decoupling introduces a length scale $\tau_d$. Pair
annihilation introduces a second scale, $\tau_a$. When
$\tau_a\gg\tau_d$ these two physical effects can be separated. Thus
we consider first the case of purely collisionless CDM, i.e.
$\gamma=0$ in (\ref{fluideqs}), in which case there is no kinetic
decoupling. Now there is only one timescale in the problem,
$\tau_a$, and we denote the corresponding transfer functions in
(\ref{delupa}) by $f_3^{\rm pa}$ and $f_4^{\rm pa}$. The results
obtained by numerical integration are shown in Fig.\
\ref{fig:figfgam0}.  For $k\tau_a\ll1$, $f_3^{\rm
pa}\approx1-3.00(k\tau_a)^4$ (where the coefficient 3.00 was found
by numerical integration) and $f_4^{\rm pa}\approx{\cal
C}-\frac{1}{2}$. For $k\tau_a\gg1$, the WKB approximation yields
\begin{eqnarray}\label{f34wkb}
  f_3^{\rm pa}&\approx&1-0.1067(k\tau_a)^{-2}\ ,\nonumber\\
  f_4^{\rm pa}&\to&-\frac{\pi}{2}\sin\alpha+
    \left[\frac{1}{C(k)}-1\right]\ln(9.82k\tau_a)\ ,
\end{eqnarray}
where the coefficients $0.1067$ and $9.82$ were determined by
numerical integration.  The errors made by assuming (\ref{f34wkb})
are less than 1\% for $k\tau_a>100$.

For WIMP dark matter, the effects of kinetic decoupling and pair
annihilation combine to give transfer functions $f_3^{\rm kd+pa}$
and $f_4^{\rm kd+pa}$ in (\ref{delupa}).  For $T/m_\chi=0$ and
$\tau_d\ll\tau_a\ll\tau$ these are
\begin{eqnarray}\label{kd+pa}
  f_3^{\rm kd+pa}(k)&=&f_1(k)f_3^{\rm pa}(k)\ ,\nonumber\\
  f_4^{\rm kd+pa}(k)&=&f_4^{\rm pa}(k)+\frac{1}{C(k)}\left[
    f_2(k)-{\cal C}+\frac{1}{2}\right]\\
  &&\!\!\!\!\!\!\!\!\!\!\!\!\!\!\!\!\!\!\!\!\!\!\!\!\!\!\!\!
    +\left[f_1(k)-1\right]\left\{\left[\frac{1}{C(k)}-1\right]
    \ln(1.589k\tau_a)-\ln C(k)\right\}\ .\nonumber
\end{eqnarray}
The coefficient $1.589$ was found numerically but otherwise
$f_3^{\rm kd+pa}$ and $f_4^{\rm kd+pa}$ were determined
analytically.  The general transfer function is given in terms of
separate transfer functions for kinetic decoupling $(f_1,f_2)$ and
pair annihilation $(f_3^{\rm pa}, f_4^{\rm pa})$.

Pair annihilation modifies the CDM velocity and density transfer
functions at $k\sim\tau_a^{-1}$ due to the peak in $f_3^{\rm
pa}(k)$. There is also a small modification of the density for
$k\tau_a\gg1$.  In the limit $k\tau\gg k\tau_a\gg1$,
\begin{equation}\label{nushort}
  \frac{\nu_0}{9}\to f_1(k)\left[\ln\theta+0.08909\ln\left(
    \frac{7.416\tau_a}{\tau}\right)\right]+f_2(k)\ .
\end{equation}
Comparing this with (\ref{coldlimsol}), we see that the term
proportional to $1-C(\infty)=0.08909$ is induced by pair
annihilation.  This effect arises from the reduction of the
gravitational potential by a factor $C(\infty)=\dot a(\tau_1)/ \dot
a(\tau_2)$ at short wavelengths after pair annihilation. However,
the correction to $\nu$ is always less than $9\%$.

The full density transfer function for nonzero CDM temperature
during the radiation-dominated era follows from (\ref{tempfit}),
(\ref{delupa}), and (\ref{kd+pa}).  Numerical tables of the transfer
functions $f_1(k\tau_d)$, $f_2(k\tau_d)$, $C(k\tau_a)$,
$\alpha(k\tau_a)$, $f_4^{\rm pa}(k\tau_a)$, and $f_4^{\rm
pa}(k\tau_a)$ are available from the author.

\section{Evolution through Radiation-Matter Equality}

Following kinetic decoupling and pair annihilation there is one
additional major life event for CDM density fluctuations before they
collapse to form nonlinear structures: the transition to a
matter-dominated universe occurring at $1+z_{\rm eq}\equiv a_{\rm
eq}^{-1}\approx3200$.  The constituents include photons, neutrinos,
baryons, and CDM.  The background equation of state is modified by
the presence of nonrelativistic baryons and CDM,
\begin{equation}\label{equality}
  3w=\frac{1}{1+y}\ ,\ \ y=\frac{\rho_b+\rho_c}{(g_{\rm eff}/2)
    \rho_\gamma}=\frac{a}{a_{\rm eq}}\ .
\end{equation}
Assuming that dark energy and spatial curvature can be neglected
during the times of interest, the solution of the Friedmann equation
for $\tau\gg\tau_a$ is
\begin{equation}\label{friedsol}
  \sqrt{1+y}=1+\frac{\tau}{2\tau_e}\ ,\ \ \tau_e\equiv
  \sqrt{\frac{a_{\rm eq}}{\Omega_m H_0^2}}\ .
\end{equation}
Here $\Omega_m$ is the present-day density parameter for
$\rho_m\equiv\rho_b+\rho_c$.

Before recombination, Thomson scattering couples the photons and
baryons so that they behave as a single fluid with sound speed given
by
\begin{equation}\label{cgammab2}
  3c_{\gamma b}^2=\left(1+\frac{3}{4}\frac{\rho_b}{\rho_\gamma}
    \right)^{-1}\ .
\end{equation}
We make two approximations which enable us to reduce the dynamics to
a fourth-order system in time, for which we obtain limiting analytic
solutions: Photons and baryons behave like a single perfect fluid
and neutrino shear stress is neglected. (We also assume a flat
background, $K=0$, but this is always a good approximation during
the times of interest.) These approximations introduce small errors
in the results which may be eliminated by integrating the full
system of equations for photons, neutrinos, baryons, and CDM with
CMBFAST \cite{cmbfast} or equivalent starting after pair
annihilation from initial conditions obtained in the preceding
section.  However, the simplified dynamics leads to analytic results
making it easy to distinguish the various physical effects in the
transfer functions.

With these assumptions and the further restrictions $\tau\gg\tau_a$
and $T_d/m_\chi\ll1$, the perturbed Einstein equations may be
combined to yield an equation of motion for the gravitational
potential $\Phi=\Psi$,
\begin{eqnarray}\label{Phimatt}
  \ddot\Phi+3(1+c_{\gamma b}^2){\cal H}\dot\Phi+3(c_{\gamma b}^2-w)
    {\cal H}^2\Phi+k^2c_{\gamma b}^2\Phi\qquad&&\nonumber\\
  =-A_cc_{\gamma b}^2\left(\delta_c-\frac{\rho_b\rho_\nu}{\rho_c\rho_\gamma}
    \delta_\nu\right)\ ,&&
\end{eqnarray}
where $\delta_c$ and $\delta_\nu$ are, respectively, the CDM and
neutrino density perturbation in conformal Newtonian gauge, and we
have defined
\begin{equation}\label{Acdef}
  A_c\equiv4\pi Ga^2\rho_c=\frac{3}{2}\frac{(1-f_b)y}{1+y}{\cal H}^2
    \ ,
\end{equation}
where $f_b\equiv\Omega_b/\Omega_m$.  Equation (\ref{Phimatt}) is a
modified form of (\ref{Phieom}). The CDM fluid equations
(\ref{fluideqs}) remain valid with $\delta=\delta_c$, $u=u_c$.

The evolution of metric perturbations is coupled to the evolution of
both CDM and neutrino perturbations.  To simplify the presentation
we avoid solving the collisionless Boltzmann equation for neutrinos.
Instead we make an additional approximation for the neutrino
dynamics: either the neutrino fluid evolves like the photon-baryon
fluid, or neutrino perturbations are damped by Hubble expansion on
sub-horizon scales in a manner consistent with their evolution on
super-horizon scales. In the first case we set
$\delta_\nu=\delta_\gamma$ on all scales, which is equivalent to
setting $\delta_\nu=0$ in Eq.\ (\ref{Phimatt}) and replacing
$c_{\gamma b}$ with $c_{rb}$ defined by
\begin{equation}\label{crb2}
  3c_{rb}^2=\left[1+\frac{3}{4}\frac{\rho_b}{(g_{\rm eff}/2)
    \rho_\gamma}\right]^{-1}=\left(1+\frac{3}{4}f_by\right)^{-1}\ .
\end{equation}
In the second case, we note that on scales much larger than the
Hubble length, for isentropic (curvature) perturbations all species
have the same number density perturbation,
$\delta_c=\delta_\nu=\delta_\gamma$. The evolution of isentropic
perturbations on large scales follows from Ref.\ \cite{b06}:
\begin{equation}\label{dells}
  \delta_\nu\approx-\left(\frac{2}{1+w}\right)(\Phi+{\cal H}^{-1}
    \dot\Phi)\ .
\end{equation}
On scales much smaller than the Hubble (or neutrino free-streaming)
length, during the radiation-dominated era $\delta_\nu$ undergoes
damped oscillations while the amplitude of the photon-baryon
oscillations is constant.  This qualitative behavior is correctly
reproduced if one applies (\ref{dells}) to the neutrinos at all
times while leaving (\ref{Phimatt}) unchanged.

Thus, our two approximations for neutrino evolution correspond to
two different choices for the sound-speed of the photon-baryon gas:
either the sound speed is set to $c_{rb}$ with $\delta_\nu=0$ or the
sound speed is $c_{\gamma b}$ with $\delta_\nu$ given by
(\ref{dells}). Either way, when the CDM dynamics is added, the
evolution reduces to a fourth-order system in time.  The difference
between the two treatments gives a measure of the importance of
massless neutrinos for CDM evolution.

Setting $\delta_\nu=0$ and writing the sound speed as $c_b$ which
may be set to either $c_{\gamma b}$ or $c_{rb}$, differentiating
(\ref{Phimatt}) and using the CDM continuity equation gives
\begin{eqnarray}\label{ucconstr}
  \partial_t^3\Phi+5{\cal H}\ddot\Phi+\frac{3}{2}{\cal H}^2
  \dot\Phi[3-5w-3c_b^2(1+w)]+3A_cc_b^2\dot\Phi\nonumber\\
  +k^2c_b^2(\dot\Phi+{\cal H}\Phi)=A_ck^2c_b^2u_c\ .\qquad
\end{eqnarray}
Differentiating this again and using the velocity equation $\dot
u_c+{\cal H}u_c=\Phi$ gives
\begin{eqnarray}\label{Phi4dot}
  \partial_t^4\Phi+(8-3c_b^2){\cal H}\partial_t^3\Phi
    +\left[17-15(w+c_b^2)-\frac{6wc_b^2}{c_{rb}^2}\right]
    {\cal H}^2\ddot\Phi&&\nonumber\\
  +\frac{9}{2}\left[(2-3c_b^2)(1-3w)+\frac{4w}{3}
    \left(1-\frac{c_b^2}{c_{rb}^2}\right)\right]
    {\cal H}^3\dot\Phi&&\nonumber\\
  +k^2c_b^2\left[\ddot\Phi+3{\cal H}\dot\Phi
    +\left(\frac{2}{c_{rb}^2}-3\right)w{\cal H}^2
    \Phi\right]=0\ .\quad&&
\end{eqnarray}
For wavelengths longer than the Hubble length this equation is
correct if $c_b=c_{rb}$; for wavelengths much shorter than the
Hubble length it is approximately correct if $c_b=c_{\gamma b}$. We
now present the analytic solution to the fourth-order system in
these two limits.

In the long-wavelength limit $k^2\ll{\cal H}^2$, the four linearly
independent solutions may be obtained as quadratures using the
methods of Ref.\ \cite{b06}:
\begin{subequations}\label{4sol0}
\begin{eqnarray}
  \Phi_1&=&\frac{{\cal H}\tau_e}{y^2}=\frac{\sqrt{1+y}}{y^3}
    \ ,\label{Phi01}\\
  \Phi_2&=&\frac{3{\cal H}}{2y^2}\int^\tau y^2(1+w)\,d\tau
    \ ,\label{Phi02}\\
  \Phi_3&=&\frac{{\cal H}}{y^2}\int^\tau y^3w\,d\tau
    \ ,\label{Phi03}\\
  \Phi_4&=&\frac{{\cal H}}{y^2}\int^\tau y^2w
    [1+(\rho_b/\rho_m)y]\,d\tau\ .\label{Phi04}
\end{eqnarray}
\end{subequations}
These solutions give the time-dependence in the limit $k\to0$ valid
in both the radiation- and matter-dominated eras. The physical
solution for isentropic curvature fluctuations is
\begin{eqnarray}
  \Phi_+&=&\frac{3}{2}\Phi_2+\frac{8}{5}\Phi_1\nonumber\\
  &=&\frac{9}{10}\left(1+\frac{2}{9y}-\frac{8}{9y^2}-\frac{16}{9y^3}
    +\frac{16\sqrt{1+y}}{9y^3}\right)\nonumber\\
  &\approx&1-\frac{1}{16}y\ \ \hbox{for}\ \ y\ll1\ .
    \quad
\end{eqnarray}
Mode 1 is the decaying mode.  Modes 3 and 4 are entropy
perturbations and (for $k^2\ll{\cal H}^2$) are constants added to
the other solutions,
\begin{equation}\label{entsol0}
  \Phi_3=\frac{4}{3}-2\Phi_2\ ,\ \ \Phi_4=-1+\frac{5}{3}\Phi_2
    +\frac{\rho_b}{\rho_m}\Phi_3\ .
\end{equation}

In the opposite limit, $k^2\gg{\cal H}^2$, approximate solutions to
(\ref{Phi4dot}) may be found using the WKB method. The first-order
WKB approximation gives two oscillatory solutions,
\begin{equation}\label{wkbosc}
  \Phi_\pm=(a^2c_b^{3/2})^{-1}\exp\pm ik\int^\tau c_b\,d\tau\ ,
\end{equation}
and two non-oscillatory solutions of the equation
\begin{equation}\label{denonosc}
  \ddot\Phi+3{\cal H}\dot\Phi+\left(\frac{2}{c_{rb}^2}-3\right)
    w{\cal H}^2\Phi=0\ .
\end{equation}
This equation is equivalent to the usual evolution equation for cold
dark matter perturbations calculated assuming that the other
components are unperturbed:
\begin{equation}\label{Dgrowth}
  \ddot D+{\cal H}\dot D=A_cD\ ,\ \ D\propto\delta\propto y\Phi\ .
\end{equation}
For $a(\tau)$ given by (\ref{friedsol}), the exact solutions of this
equation are \cite{ghs05}
\begin{eqnarray}\label{Dsol}
  &&D_+(\tau)=P_\nu(1+\tau/2\tau_e)\ ,\ \
  D_-(\tau)=Q_\nu(1+\tau/2\tau_e)\ ,\nonumber\\
  &&\qquad\qquad\qquad
  \nu\equiv-\frac{1}{2}+\frac{1}{2}\sqrt{25-24f_b}\ ,
\end{eqnarray}
where $P_\nu$ and $Q_\nu$ are Legendre functions of the first and
second kind. Early in the radiation-dominated era they give
\begin{equation}\label{legfun}
  D_+\approx1+\frac{\nu(\nu+1)}{4}\frac{\tau}{\tau_e}\ ,\ \
  D_-\approx\frac{1}{2}\ln\left(\frac{4\tau_e}{\tau}\right)
  \ \ \hbox{for $\tau\ll\tau_e$}\ .
\end{equation}
In the opposite limit,
\begin{eqnarray}\label{Dsolbig}
  D_+&\approx&\frac{\Gamma(\nu+\frac{1}{2})}
    {\sqrt{\pi}\,\Gamma(\nu+1)}\left(\frac{\tau}{\tau_e}\right)^\nu
  \nonumber\\
  &\approx&\left[\left(\nu+\frac{1}{2}\right)\frac{\tau}{\tau_e}
    D_-\right]^{-1}\ \ \hbox{for $\tau\gg\tau_e$}\ .
\end{eqnarray}

By matching the solutions found here to those in the
radiation-dominated era, (\ref{Phipa}) and (\ref{delupa}), we obtain
results for the gravitational potential $\Phi$ and synchronous gauge
density perturbation $\nu$ (not to be confused with the index of the
Legendre functions) valid through the matter-radiation transition.
On large scales,
\begin{equation}\label{longeq}
  \Phi\approx-\Phi_+\ ,\ \ \nu\approx\frac{2(k\tau_e)^2y^2\Phi_+}
    {4+3y}\ ,\ \ k^2\tau_e^2\ll1\ ,
\end{equation}
where we have chosen the normalization $\Phi=-1$ as $k\tau\to0$.  On
scales shorter than the Jeans length,
\begin{eqnarray}\label{shorteq}
  \frac{\nu}{9Cf_3}&\approx&D_-+D_+\ln\left(\frac{4k\tau_e}{\sqrt{3}}
    \right)\ ,\nonumber\\
  \Phi&\approx&-\frac{3C(c_b\sqrt{3})^{1/2}}{(kc_b\tau_e)^2}
  \frac{\cos(\theta+\alpha)}{y^2}-\frac{3(1-f_b)\nu}{2(k\tau_e)^2y}
    \ ,\qquad
\end{eqnarray}
where $C\equiv C(k)$ and $f_3\equiv f_3^{\rm kd+pa}(k)$ were given
in the previous section.  For $k\ll\tau_a^{-1}$, $C=f_3=1$.

\begin{figure}[t]
  \begin{center}
    \includegraphics[scale=0.65]{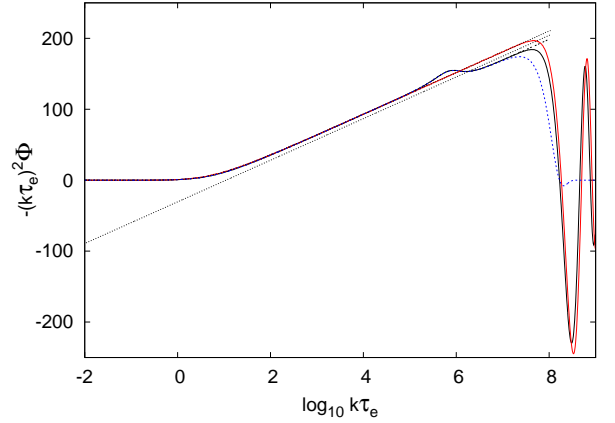}
  \end{center}
  \caption{Results for the gravitational potential transfer function
    at $y=a/a_{\rm eq}=10$ after the radiation-matter transition,
    assuming $T_d=22.6$ MeV, $\Omega_mh^2=0.131$, $\Omega_bh^2=0.022$.
    The solid black curve shows the effects of pair annihilation at
    $T\approx0.5$ MeV as the bump at $k\tau_e\approx10^{5.8}$ and
    the effects of kinetic decoupling (with $T_d/m_\chi=0$) as the
    damped oscillations at $k\tau_e>10^7$; the rapidly damped
    dotted curve includes the effects of free-streaming damping
    assuming $(2T_d/m_\chi)^{1/2}=0.0213$.  Other curves show the
    effect of eliminating either pair annihilation or kinetic
    decoupling.  The straight dotted line is the WKB solution of
    Eq.\ (\ref{shorteq}), vertically offset for clarity.}
  \label{fig:figkd+pa+eq}
\end{figure}

Figure \ref{fig:figkd+pa+eq} shows the results obtained from
numerical integration of (\ref{Phimatt}) using (\ref{dells}) to
approximate the neutrino evolution.  In order to illustrate the
logarithmic wavenumber-dependence of the growing mode in the
matter-dominated era, the potential is scaled by $-(k\tau_e)^2$ and
is plotted in a semilog fashion.  If instead of using (\ref{dells})
the neutrinos are assumed to evolve like the photon-baryon fluid,
the maximal change in $\Phi$ is about 4\% occurring at
$k\tau_e\approx2$ with negligible differences at much higher and
lower frequencies.  Thus the CDM transfer function is relatively
insensitive to the detailed dynamics of massless neutrinos.  For the
CDM and baryon abundances used in the calculations, $D_+=11.34$ at
$y=10$.

The WKB result helps explain how gravitational potential
fluctuations with an rms amplitude $2\times10^{-5}$ on large scales
can generate nonlinear structures at high redshift. Growth after the
universe becomes matter-dominated contributes a factor
$D_+\approx1+\frac{3}{2}y$ once the baryons are released and join
the CDM potential wells. Evolution after kinetic decoupling in the
early universe contributes an additional factor
$9\ln(4k\tau_e/\sqrt{3})$.  The logarithm is sometimes called the
Meszaros effect \cite{mesz}; it arises because acoustic oscillations
in the relativistic plasma gravitationally induce velocity
perturbations in the dark matter.  Half of the factor of 9 (i.e., a
factor of 3) arises because the comoving number density is affected
by the cube of the strain $(1-\Psi)$ in three dimensions.
Altogether, CDM density perturbations in the matter-dominated era
are enhanced by a factor $\frac{27}{2}y\ln(4k\tau_e/\sqrt{3})$ which
is sufficient to drive rms density perturbations nonlinear on scales
below $10^{-5}$ M$_\odot$ by a redshift of 20.

These results neglect free-streaming of the CDM particles.  As shown
in Section \ref{sec:fluid}, during the radiation-dominated era CDM
free-streaming leads to a modified Gaussian suppression, Eq.\
(\ref{tempfit}).  Including the radiation-matter transition, this
gives
\begin{equation}\label{Lfs}
  (k_{\rm fs}\tau_d)^{-1}=\sqrt{\frac{6T_d}{5m_\chi}}\ln\left(
    \frac{1+4\tau_e/\tau_\ast}{1+4\tau_e/\tau}\right)\ ,
\end{equation}
where $\tau_\ast\approx\tau_d$.  For the parameters of Fig.\
\ref{fig:figkd+pa+eq} with pair annihilation,
$\tau_e/\tau_d=3.895\times10^7$.

\section{Smallest-scale CDM structures}

Having computed the evolution of the CDM transfer function through
kinetic decoupling, pair annihilation, and radiation-matter
equality, we can now make predictions for the smallest-scale CDM
structures.  To be definite, we adopt the standard flat $\Lambda$CDM
model with $\Omega_bh^2=0.022$, $\Omega_ch^2=0.109$, $h=0.71$,
$\Omega_\Lambda=0.740$.  With these assumptions, $a_{\rm
eq}^{-1}=3160$, $\tau_e=147$ Mpc, $\tau_d=3.78$ pc, and $k_{\rm
fs}^{-1}=1.18$ pc.

The CDM density transfer function $\nu(k,\tau)$ is normalized so
that the variance of $\delta\rho/\rho$ in the usual synchronous
gauge is $\sigma^2=\int\nu^2(k,\tau)P_i(k)\,d^3k$ where $P_i(k)$ is
the spectral density of the gravitational potential $\Phi$ early in
the radiation-dominated era.  The radiation-era potential is related
to the curvature perturbation in a flat universe $\zeta={\cal
R}=\kappa$ \cite{b06} by $\Phi=\frac{2}{3}\kappa$.

The treatment of this paper makes several approximations: neutrino
shear stress is neglected; the Boltzmann equations for neutrinos and
photons after recombination have not been integrated; and photons
and baryons are assumed to be tightly coupled until $y=a/a_{\rm
eq}=10$ ($z=315$).  To test this crude treatment of multiple fluids,
we compared the resulting transfer function with that computed
numerically from CMBFAST \cite{cmbfast}.  For $k\tau_e\approx100$
(small enough so that pair annihilation and kinetic decoupling are
unimportant), our $\nu(k)$ was too small by 7\% (due to the delay in
baryons joining CDM until $z=315$) while for $k\tau_e<0.1$ our
$\nu(k)$ was too large by 3.6\% (due to the neglect of neutrino and
photon shear stress).  A more accurate calculation would incorporate
our transfer functions for pair annihilation and kinetic decoupling
into CMBFAST or equivalent code, but this level of accuracy is
unnecessary for the present purposes.

\begin{figure}[t]
  \begin{center}
    \includegraphics[scale=0.65]{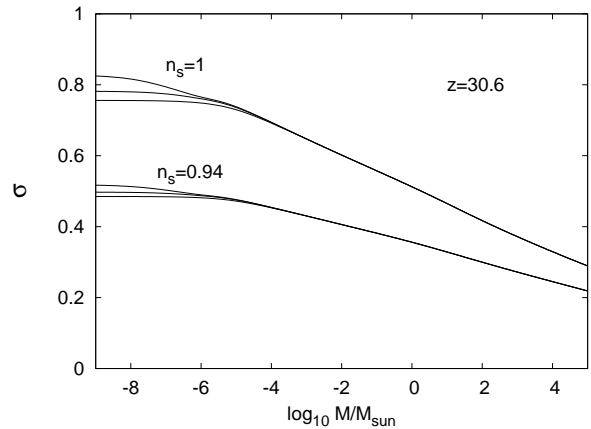}
  \end{center}
  \caption{RMS mass density perturbation in a sphere containing mass $M$,
    at redshift $z=30.6$ ($a/a_{\rm eq}=100$).  Two different
    assumptions are made for the scalar spectral slope $n_s$.  For each
    case, three different choices are shown for the neutralino thermal
    speed at decoupling: $(2T_d/m_\chi)^{1/2}\in\{0,0.01,0.0213\}$.}
  \label{fig:figsig}
\end{figure}

Figure \ref{fig:figsig} shows the rms mass density perturbation in a
sphere containing mean mass $M$, which follows from
\begin{equation}\label{massvar}
  \sigma^2(M)=\int\nu^2(k,\tau)W^2(kR)P_i(k)\,d^3k\ ,
\end{equation}
where $W(x)=3j_1(x)/x$ is the spherical tophat window function and
$R=(3M/4\pi\rho_m)^{1/3}$.  The primordial spectrum has been
normalized to $\Delta_{\cal R}^2=2.40\times10^{-9}$ at
$k=0.002/\hbox{Mpc}$ from the WMAP 3-year results \cite{spergel}.
The amplitude for the scale-invariant case $n_s=1$ is more than 50\%
greater than it is in the tilted model $n_s=0.94$ favored by the
data. With a neutralino mass of 100 GeV and kinetic decoupling
temperature 22.6 MeV, $(2T_d/m_\chi)^{1/2}=0.0213$.  In this case,
$3\sigma$ perturbations reach $\nu=1$ by $z=47$.

\begin{figure}[t]
  \begin{center}
    \includegraphics[scale=0.65]{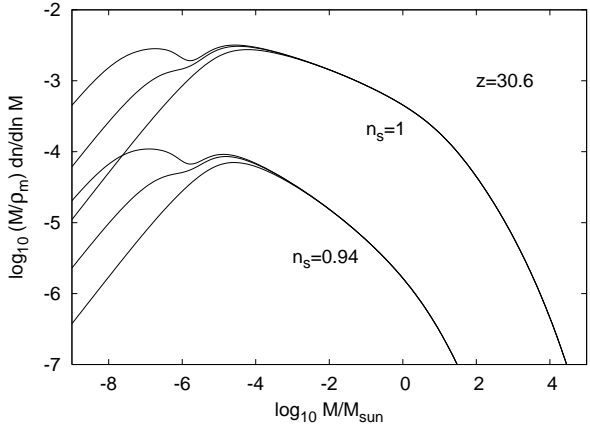}
  \end{center}
  \caption{Press-Schechter mass fraction for the cases shown in
    Fig.\ \ref{fig:figsig}.  The effects of increasing free-streaming
    damping are apparent in the erasure of the smallest objects with
    increasing $(2T_d/m_\chi)^{1/2}$.  The cases of no free-streaming
    damping have bimodal distributions, however this requires
    unreasonable assumptions about the dark matter mass and couplings.}
  \label{fig:figps1}
\end{figure}

The Press-Schechter \cite{ps} mass function $dn/dM$ obeys
\begin{equation}\label{psmf}
  \frac{M}{\rho_m}\frac{dn}{d\ln M}=\frac{\delta_c}{\sigma}
    \left(\frac{2}{\pi}\right)^{1/2}\left\vert\frac{d\ln\sigma}
    {d\ln M}\right\vert e^{-\delta_c^2/2\sigma^2}\ ,
\end{equation}
where $\delta_c=1.686$. The results, shown in Fig.\
\ref{fig:figps1}, suggest that the smallest CDM objects may have
masses even smaller than an earth mass, $3.0\times10^{-6}$
M$_{\odot}$.

WIMP decoupling imprints two different length scales on the
spectrum: the comoving horizon size $\tau_d\propto T_d^{-1}$ and the
comoving free-streaming distance $(T_d/m_\chi)^{1/2}\tau_d$; the
former scale is larger and more important. The characteristic mass
for the smallest objects is the CDM mass in the horizon at
decoupling,
\begin{eqnarray}\label{mmin}
  M_d&=&\frac{4\pi}{3}\rho_m(a=1)(c\tau_d)^3\nonumber\\
  &=&2.05\times10^{-4}\,\Omega_mh^2\left(\frac{T_d\sqrt{g_{\rm
  eff}}}{\hbox{50 MeV}}\right)^{-3}\,{\rm M}_\odot\nonumber\\
  &=&7.59\times10^{-3}\,C^{3/4}\left(\frac{m_\chi\sqrt{g_{\rm eff}}}
    {\hbox{100 GeV}}\right)^{-15/4}\,{\rm M}_\odot\ ,\qquad
\end{eqnarray}
where $C$ was defined in (\ref{Cneut}).  With the default values
$C=0.0433$ and $g_{\rm eff}=10.75$, $M_d=8.29\times10^{-6}$
M$_\odot$, or
\begin{equation}\label{Mdec}
  M_d=1.0\times10^{-4}\left(\frac{T_d}{\hbox{10 MeV}}\right)^{-3}
    \,{\rm M}_\odot\ ,
\end{equation}
in agreement with Ref.\ \cite{lz05}. Berezinsky et al.\ \cite{bde03}
derived a different characteristic mass assuming that diffusion
followed by free-streaming damping sets the minimum object mass. As
we have shown, for $(2T_d/m_\chi)^{1/2}\ll1$, the dominant damping
process is not diffusion or free streaming but friction between
WIMPs and relativistic leptons during kinetic decoupling (Silk
damping). That this is not a diffusive process may be seen by its
persistence in the limit $T_d/m_\chi=0$.

The evolution of the mass function requires computing the density
perturbations at smaller redshift. The baryons present a slight
complication because after recombination, even though they are free
to move apart from the photon gas, they resist gravitational
instabilities for $k>k_{\rm Jb}$ where the baryon Jeans wavenumber
is
\begin{equation}\label{barjeans}
  k_{\rm Jb}=\left(\frac{3\Omega_m}{2a}\right)^{1/2}\frac{H_0}{c_b}\ ,
\end{equation}
with $c_b$ being the baryon sound speed.  The small residual
ionization fraction persisting after the nominal recombination
maintains the baryon temperature close to the microwave background
temperature until $a\approx a_{\rm Tdec}=1/125$ yielding (for the
standard cosmological parameters) $k_{\rm Jb}=252$ Mpc$^{-1}$ for
$a\ll a_{\rm Tdec}$ and $k_{\rm Jb}=252\sqrt{a/a_{\rm Tdec}}$
Mpc$^{-1}$ for $a\gg a_{\rm Tdec}$. Thus, even before reionization,
baryon perturbation growth is inhibited on scales $k>252$ Mpc$^{-1}$
or dark matter masses less than about $10^5$ M$_\odot$.  Since we
are mainly interested here in smaller scales, we will assume that
baryons do not participate in gravitational instability.  The CDM
density perturbations therefore grow with time according to the
$D_+(\tau)$ solution (\ref{Dsol}) with $f_b=\Omega_b/\Omega_m$. For
$f_b=0.168$, at $a=1$ this results in an almost 50\% reduction in
linear growth compared with $f_b=0$. Dark energy plays a role only
at very late times.  At $z=2.16$, a cosmological constant with
$\Omega_\Lambda=0.74$ suppresses the growth of linear density
perturbations by 2\%.

\begin{figure}[t]
  \begin{center}
    \includegraphics[scale=0.65]{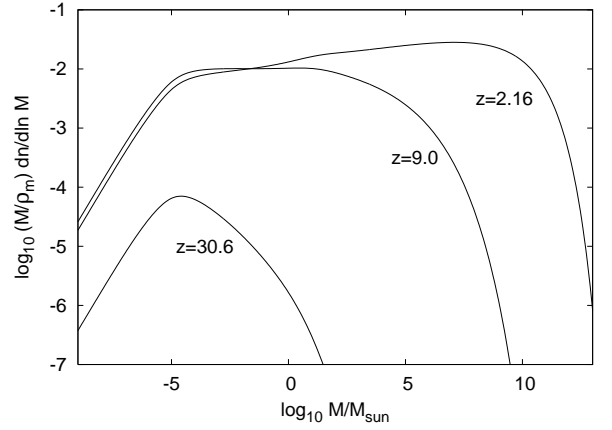}
  \end{center}
  \caption{Press-Schechter mass fraction for $n_s=0.94$,
  $m_\chi=100$ GeV and $T_d=22.6$ MeV, for several redshifts.
  Constant $(M/\rho_m)dn/d\ln M$ indicates equal mass is contained
  in objects whose masses span equal logarithmic mass intervals.
  The slight bump at 50 M$_\odot$ is due to enhanced growth
  during electron pair annihilation.  The growth on scales $M>10^5$
  M$_\odot$ has been underestimated by ignoring the boost given
  by infalling baryons.}
  \label{fig:figps2}
\end{figure}

Figure \ref{fig:figps2} shows the evolution of the mass fraction
with redshift for plausible WIMP parameters. At $z=31$ the peak of
$Mdn/d\ln M$ occurs at $M\approx 2.3M_d$ (6 earth masses) but there
may be numerous smaller objects. Indeed, the numerical results imply
$d\sigma/d\ln M\propto-(M/M_d)^{2/3}$ for $M\ll M_d$. This follows
from $W(kR)$ being an analytic function of $(kR)^2$ with $W(0)=1$.
For $R=0$ the variance integral (\ref{massvar}) converges because of
the small-scale damping caused by kinetic decoupling. Thus,
analyticity implies $\sigma^2\propto C_1-C_2(R/\tau_d)^2$ as $R\to0$
where $C_1$ and $C_2$ are independent of filter radius $R\propto
M^{1/3}$, hence $d\sigma/d\ln M\propto-(R/\tau_d)^2$ as $R\to0$, or
\begin{equation}\label{smallM}
  \frac{dn}{d\ln M}\propto M^{-1/3}\ .
\end{equation}
Given a smooth, symmetric window function, the number of small
objects diverges but the mass contained in them converges.

These results for $M<M_d$ are uncertain because there are window
functions $W(kR)$ that give a different result. For example, a sharp
$k$-space filter, with $W(kR)=1$ for $kR<\pi$ and zero otherwise,
when combined with the free-streaming damping of $\nu(k,\tau)$,
leads to an exponential cutoff instead of a power-law rise of
$dn/d\ln M$ as $M\to0$.  However, a sharp $k$-space window gives
rise to long-range oscillations in physical space, which seems
unlikely to properly describe the local physics of dark matter halo
formation.

The choice of window function represents an enormous simplification
of the nonlinear dynamics of dark matter halo formation.  N-body
simulations are needed to determine the correct mass function for
$M<M_d$.  Such simulations must fully resolve scales below the
free-streaming damping length of 1.17 pc in the initial mass
distribution.  Cold dark matter caustics \cite{caustics} are also
likely to be important in the formation and dynamics of the smallest
halos.

By redshift 9, RMS density perturbations become nonlinear and the
small-scale mass distribution is nearly scale-invariant over 8
orders of magnitude of mass, $10^{-5}$ M$_\odot$ to $10^3$
M$_\odot$, reflecting the nearly scale-invariant ($n\approx-3$)
character of the small-scale CDM density field. At later times
larger objects build up by mergers of substructures as the effective
spectral slope $n>-3$ on larger scales.  The smallest-scale
structure forms by almost sudden collapse as opposed to hierarchical
clustering.

The results presented here do not include tidal destruction of small
objects formed at high redshift.  The survivability of earth-mass
halos remains an open question \cite{diemand,bde06,gg06}. The vast
dynamic range of masses reflected in Fig.\ \ref{fig:figps2} provides
an enormous challenge to numerical simulation methods attempting to
determine the survival of the first forming halos \cite{bagla}.

\bigskip
\section{Conclusions}

The small-scale transfer function of cold dark matter encodes rich
information about WIMP physics.  Were it possible to measure the
power spectrum on scales below a solar mass, with precision similar
to cosmic microwave background measurements on degree angular
scales, we could hope to determine the WIMP mass and elastic
scattering cross section with leptons by astrophysical means.  This
seems far-fetched because of the tremendous degree of nonlinear
processing that has occurred since the first cold dark matter
structures formed at a redshift about 50.  However, if any of the
smallest bound objects survive tidal stripping, they may produce an
observable WIMP annihilation signal \cite{diemand} which provides a
window into particle physics.

Moreover, the nearly scale-invariant structure predicted for dark
matter fluctuations on mass scales below $10^4$ M$_\odot$ could
affect the formation of larger-mass structures.  This mass range is
very difficult to study with standard numerical simulation methods.

While the present work has answered the question of how small-scale
fluctuations in the WIMP distribution evolve during the linear stage
of evolution, it raises new questions about nonlinear halo
formation. If the Press-Schechter formalism with a local spatial
filter (or analytic window function) correctly describes the
formation of the smallest halos, then there is no minimum mass for
WIMP microhalos; the mass function $dn/d\ln M\propto M^{-1/3}$ for
$M<10^{-4}(T_d/\hbox{10 MeV})^{-3}$ M$_\odot$.  Most of the mass,
however, is in larger clumps.  Nonlinear evolution is likely to
strongly modify the Press-Schechter distribution.  Not only is tidal
stripping important, but the Press-Schechter model is based on the
assumption of hierarchical clustering, which breaks down on the
small scales considered here.

The present paper has  considered dark matter only in the form of a
thermal relic, i.e. particles that were once in thermal equilibrium
with the relativistic plasma of the early universe. The dark matter
may be instead a nonthermal relic --- the axion --- which began its
life as a Bose-Einstein condensate. The axion field begins
oscillating after the QCD phase transition but is never
collisionally coupled to the plasma.  It would be interesting to
investigate the smallest-scale structure in an axion dark matter
model.

\acknowledgments

I thank Jon Arons for suggesting the Sonine polynomial expansion and
Dominik Schwarz for correcting an error in the damping rate
calculation. This work was supported by the Kavli Foundation and by
NSF grant AST-0407050.

\end{document}